\documentclass[12pt,aps,prd,onecolumn,nofootinbib,superscriptaddress]{revtex4}

\usepackage{latexsym}
\usepackage{amssymb}
\usepackage{epsfig,amsmath,graphics}

\newcommand{\gsim}{\lower.7ex\hbox{$\;\stackrel{\textstyle>}{\sim}\;$}}
\newcommand{\lsim}{\lower.7ex\hbox{$\;\stackrel{\textstyle<}{\sim}\;$}}
\newcommand{\OO}{\mathcal{O}}

\newcommand{\GeV}{\text{ GeV}}

\newcommand{\half}{\frac{1}{2}}
\newcommand{\DO}{{\mbox{DO\hspace{-0.11in}$\not$\hspace{0.16in}}}}
\newcommand{\wino}{$\widetilde{W}\text{ }$}

\newcommand{\bino}{$\widetilde{B}\text{ }$}

\newcommand{\MET}{\mbox{$E_T\hspace{-0.23in}\not\hspace{0.18in}$}}
\newcommand{\SmMET}{E_T\hspace{-0.230in}\not\hspace{0.18in}}

\begin{document}

\pagestyle{plain}

\title{
\begin{flushright}
\mbox{\normalsize SLAC-PUB-13405}
\end{flushright}
\vskip 15 pt

Model-Independent Jets plus Missing Energy Searches}

\author{Johan Alwall, My-Phuong Le, Mariangela Lisanti and Jay G. Wacker}
\affiliation{
Stanford Linear Accelerator Center, Menlo Park, CA 94025\\
\mbox{Stanford Institute for Theoretical Physics, Stanford University,
Stanford, CA 94305}
\linebreak
\linebreak
\linebreak
}

\begin{abstract}
\begin{center}
\textbf{Abstract}
\end{center}
We present a proposal for performing model-independent jets plus missing energy searches.  Currently, these searches are optimized for mSUGRA and are consequently not sensitive to all kinematically-accessible regions of parameter space.  We show that the reach of these searches can be broadened by setting limits on the differential cross section as a function of the total visible energy and the missing energy.  
These measurements only require knowledge of the relevant Standard Model backgrounds and can be subsequently used to limit any theoretical model of new physics.  We apply this approach to an example where gluinos are pair-produced and decay to the LSP through a single-step cascade, and show how sensitivity to different gluino masses is altered by the presence of the decay chain.  The analysis is closely based upon the current searches done at the Tevatron and our proposal requires only small modifications to the existing techniques.  We find that within the MSSM, the gluino can be as light as 125 \GeV.  The same techniques are applicable to jets and missing energy searches at the Large Hadron Collider.     
\end{abstract}

 \maketitle
\newpage

\section{Introduction}

One of the most promising signatures for new physics at hadron colliders are events with jets and large missing transverse energy ($\MET$).  These searches are very general and cover a wide breadth of potential new theories beyond the Standard Model.  
%
%
Jets + $\MET$ searches pose a significant challenge, however, because the Standard Model background is difficult to calculate in this purely hadronic state.  
The general nature of the signature motivates performing a search that only requires
calculating the Standard Model background.  The challenge, then, is to minimize the risk of missing new physics while still
accounting for our limited understanding of the background.
All experimental searches of jets + $\MET$ at hadron colliders have been model-dependent, attempting to be sensitive to specific models \cite{Alitti:1989ux,Portell:2006qb, Abazov:2007ww,CDF,MonojetSearches,Mrenna:1995ax}.  Initial studies for the Large Hadron Collider (LHC) have been dominantly model-dependent \cite{Baer:1995nq,Hinchliffe:1996iu,Gjelsten:2004ki,Gjelsten:2005aw}.  In this article, we explore how modest modifications to the existing  jets and $\MET$ studies can allow them to be model-independent, broadening the reach of the experimental results in constraining theoretical models.   

Currently, jets plus $\MET$ searches at the Tevatron are based on the Minimal Supersymmetric Standard Model (MSSM) \cite{Dimopoulos:1981zb} and look for production of gluinos ($\tilde{g}$) and squarks ($\tilde{q}$), the supersymmetric partners of gluons and quarks, respectively \cite{Portell:2006qb,Abazov:2007ww,CDF}.  These particles subsequently decay into the stable, lightest supersymmetric particle (LSP), which is frequently the bino, the supersymmetric partner of the photon. The MSSM  contains hundreds of parameters and it is challenging to place mass bounds in such a multi-parameter space.  To make this tractable, the CMSSM (or mSUGRA) \emph{ansatz} has been used \cite{CMSSM}.  The CMSSM requires common scalar masses ($m_0$), gaugino masses ($m_{\half}$), and trilinear scalar soft couplings ($A_0$) at the unification scale, in addition to electroweak symmetry breaking, gauge coupling unification, and R-parity conservation.  The entire particle spectrum is determined by five parameters.

 One important consequence of this theory is that the ratio of gaugino masses is fixed at approximately
$m_{\tilde{g}} : m_{\widetilde{W}} : m_{\widetilde{B}} \simeq 6 : 2 : 1$,
where $\widetilde{W}$ refers to the triplet of winos $(\widetilde{W}^{\pm}, \widetilde{W}^0)$, the supersymmetric partners of the electroweak gauge bosons.  
Due to the number of constraints in the CMSSM, the bino is the LSP throughout the range of parameter space that the Tevatron has access to.  Furthermore, due to the renormalization group running of the squark masses, the squarks are never significantly lighter than the gluino.   Thus, the ratio in masses between the lightest colored particle and the LSP is essentially fixed.   
The CMSSM is certainly not representative of all supersymmetric models (see, for example, \cite{Dimopoulos:1996yq,Giudice:1998bp,Choi:2004sx,Kitano:2005wc,Gherghetta:1999sw,Randall:1998uk}), let alone the wider class of beyond the Standard Model theories that jets and $\MET$ searches should have sensitivity to.   Verifying that a jets and $\MET$ search has sensitivity to  the CMSSM does not mean that the search is sensitive to a more generic MSSM.   


Existing searches for gluinos and squarks make strong assumptions about the spectrum and it is unclear what the existing limits on squark-like and gluino-like particles are.  Because squarks have electric charge, LEP can place limits of $92 \GeV$ on their mass \cite{Barate:1999cn}; however, gluinos do not couple to either the photon or $Z^0$ and so limits from LEP2 are not strong.
Currently, the tightest model-independent bound on color octet fermions (such as gluinos) comes from thrust data at ALEPH \cite{Heister:2003aj} and OPAL \cite{Abbiendi:2004qz}.  New colored particles should contribute at loop-level to the running of the strong coupling constant $\alpha_s$.  To date, the theoretical uncertainties in the value of $\alpha_s$ have decreased its sensitivity to new particle thresholds.  Advances in Soft-Collinear Effective Theory, however, have been used to significantly reduce the uncertainties in $\alpha_s$ from LEP data.  The current bound on color octet fermions is 51.0$\GeV$ at 95\% confidence \cite{Kaplan:2008pt}; no limit can be set for scalar color octets.  

There is no unique leading candidate for physics beyond the Standard Model; therefore, searches for new physics need to be performed in many different channels.  Ideally, one should perform totally model-independent searches that only employ the Standard Model production cross section for physics with the desired channels and the correct kinematics.  The goal is to be sensitive to a large number of different models at the same time so that effort is not wasted in excluding the same parts of Standard Model phase space multiple times. 

Some progress on experimental model-independent searches has been made.  In an ambitious program, the CDF Collaboration at the Tevatron has looked at all possible new channels simultaneously (i.e., Vista, Sleuth, Bumphunter) \cite{Aaltonen:2007kp, Aaltonen:2007dg,Henderson:2008ps}; however, these searches have some drawbacks over more traditional, channel-specific searches.  The most important drawback is that it is difficult, in the absence of a discovery, to determine what parts of a given model's parameter space are excluded.

On the theoretical front, MARMOSET \cite{ArkaniHamed:2007fw} is a hybrid philosophy that attempts to bridge model-independent and model-dependent searches with the use of On-Shell Effective Theories (OSETs).  OSETs parameterize the most experimentally relevant details of a given model -- i.e., the particle content, the masses of the particles, and the branching ratios of the decays.  By using an on-shell effective theory, it is possible to easily search through all experimentally relevant parameters quickly.  The on-shell approximation is not applicable in all situations, but OSETs can still give a rough idea of where new physics lies.

In this article, we will explore the discovery potential of jets and missing energy channels.  In previous work \cite{Alwall:2008ve}, we presented a simple effective field theory that can be used to set limits on the most relevant parameters for jets and missing energy searches: the masses of the particles.  While this approach seems obvious, existing searches at hadron colliders (Tevatron Run II, Tevatron Run I, UA2, UA1) are based on CMSSM-parameterized supersymmetry breaking.  The previous paper studied how varying the decay kinematics changed the sensitivity of the searches and pointed out regions of parameter space where sensitivity is particularly low due to kinematics.  However, this gluino-bino module was still a model-dependent analysis in that it assumed pair-production of a new colored fermionic particle directly decaying to a fermionic LSP.

This paper will extend the analysis in two ways.  First, we propose a completely model-independent analysis for jets and missing energy searches.  This approach only requires knowledge of the Standard Model and places limits on differential cross sections, from which it is possible to set model-dependent limits.  In the second portion of the paper, we use this approach to extend our previous analysis of a directly decaying colored particle to contain a single-step cascade and study how this altered spectrum affects the final limits on the gluino's mass.

\section{Overview of Models}
\label{Sec: models}

Before continuing with the main theme of the article, let us take a moment to describe the class of models that jets + $\MET$ searches are sensitive to.  There are two general classes of particle spectra that will be covered by such searches, each of which has a stable neutral particle at the bottom of the spectrum.  Typically, the stability of these neutral particles is protected by a discrete symmetry (e.g., R-parity, T-parity, or KK-parity) and, consequently, these particles are good candidates for the dark matter.  In one class of models, the theory contains a new colored particle that cascade decays into the dark matter.  In the other class, new electroweak gauge bosons are produced.  The dark matter particle may either be produced along with the new bosons, or may be the final step in their decays.

The first class can be thought of as being generally SUSY-like where the lightest colored particle is dominantly produced through
the Standard Model's strong force.  The lightest colored particle then cascade decays down to the stable, neutral particle at the bottom of that sector.  These cascades will either be lepton-poor or lepton-rich.  Lepton-poor cascades
occur when there is no state accessible in the cascades that have explicit lepton number (e.g., sleptons)
and frequently occur when the cascades are mediated by $W^\pm$, $Z^0$, or Higgs bosons.   A simple supersymmetric example
of a lepton-poor cascade decay is a theory where the scalar masses are made heavy and only gauginos and Higgsinos are available in the decay chains.   This occurs, for instance, in PeV supersymmetry models, where the scalars are around 1000 TeV and the  fermions of the MSSM are in the 100 GeV to 1 TeV range.  Producing the color-neutral states of such a theory is difficult at hadron machines; consequently, the production of new particles will occur primarily through the decay of the gluino.

One potential cascade decay of the gluino, which will be considered in further detail in the second half of the paper, is 
\begin{equation}
\tilde{g}  \rightarrow \bar{q}_1 q_2  \widetilde{W} \rightarrow \bar{q}_1 q_2 \bar{q}_3 q_4 \widetilde{B}.
\label{eq: decay}
\end{equation}
In this cascade, the \wino decays directly into the \bino and a $W^{\pm},Z^0$ boson, which subsequently decays to two jets.  This single-step decay is the dominant cascade if the gaugino masses are unified at high energies; in this case, the branching ratio of the gluino into the wino is $\sim 80\%$.  While these cascade decays are to some degree representative, the precise mass ratio of $m_{\tilde{g}}: m_{\widetilde{W}}:m_{\widetilde{B}}$ makes a significant difference in  the searches.  In the limit where $m_{\widetilde{W}} \rightarrow m_{\widetilde{B}}$ the energy from $\bar{q}_3$ and $q_4$ is small, while if $m_{\widetilde{W}} \rightarrow m_{\tilde{g}}$  the jets from $\bar{q}_1$ and $q_2$ are soft.  If $m_{\widetilde{W}} >m_{\tilde{g}}$, this cascade is forbidden.   Interestingly, spectra with unified gaugino masses are the most difficult to see because all four jets are fairly hard and diminish the missing energy in the event in comparison to the direct decay of the gluino, $\tilde{g}  \rightarrow \bar{q}_1 q_2  \widetilde{B}.$

Leptons from the decay of the $W^{\pm}$, $Z^0$ boson can be used in the analysis as well (see Sec. \ref{Sec: leptons}).  However, jets + $\MET$ + lepton studies are better suited for lepton-rich cascades.  The addition of leptons to the searches makes the experimental systematics easier to control and improves trigger efficiencies.  Not all spectra of new physics can be probed with  these types of searches, though, and they are thus complimentary to the jets + $\MET$ search.

Other cascades may produce a greater number of jets as compared to (\ref{eq: decay}).  In NMSSM theories where there is a new singlino at the bottom of the spectrum \cite{Balazs:2007pf}, it is possible to have cascade decays that start with the gluino, go to wino plus two jets, then bino plus two additional jets, and conclude with the singlino plus two more jets.  The additional step in the decay process further diminishes the amount of  missing energy in typical events, resulting in reduced limits on spectra.  Other models, such as Universal Extra Dimensions (UEDs) \cite{Appelquist:2000nn} and Little Higgs models with T-parity \cite{ArkaniHamed:2002qx} also have new colored particles that subsequently cascade decay.  The details of the exact spectra can alter the signal significantly as jets can become soft and missing energy is turned into visible energy.   

It is also possible that new electroweak gauge bosons are produced, which then cascade decay, producing jets before ending with the neutral stable particle.  
Little Higgs models with T-parity are one such example.  In such models, the new heavy bosons $W_H^{\pm}$ and $Z_H^0$ are produced through s-channel processes.  The $W_H^{\pm}$ can decay to the $W^{\pm}$ and the dark matter $A_H$, while the $Z_H^0$ can decay to the $A_H$ and higgs.  It is also possible to produce the $W_H^{\pm}$ directly with the $A_H$ through an s-channel $W^{\pm}$ boson.  This vertex, however, is suppressed in comparison to the other two.  



\section{Proposed Analysis Strategy}
\label{Sec: proposal}

At the Tevatron, the jets + $\MET$ channel is divided into four separate searches (monojet, dijet, threejet, and multijet), with each search defined by jet cuts $\OO(30 \GeV)$.  Cuts on the missing transverse energy and total visible energy\footnote{The total visible energy $H_T$ is defined as the scalar sum of the transverse momenta of each jet.} $H_T$ of each event take place during the final round of selection cuts.  The $\MET$ and $H_T$ cuts are optimized for ``representative" points in CMSSM parameter space for each of the (inclusive) $1j - 4^+j$ searches.  However, these $\MET$ and $H_T$ cuts may not be appropriate for theories other than the CMSSM.  Indeed, considering the full range of kinematically allowed phase space means accounting for many combinations of missing and visible energy.  A set of static cuts on $\MET$ and $H_T$ is overly-restrictive and excludes regions of phase space that are kinematically allowed.

\begin{figure}[t,b] 
   \centering
     \includegraphics[width=3.2in]{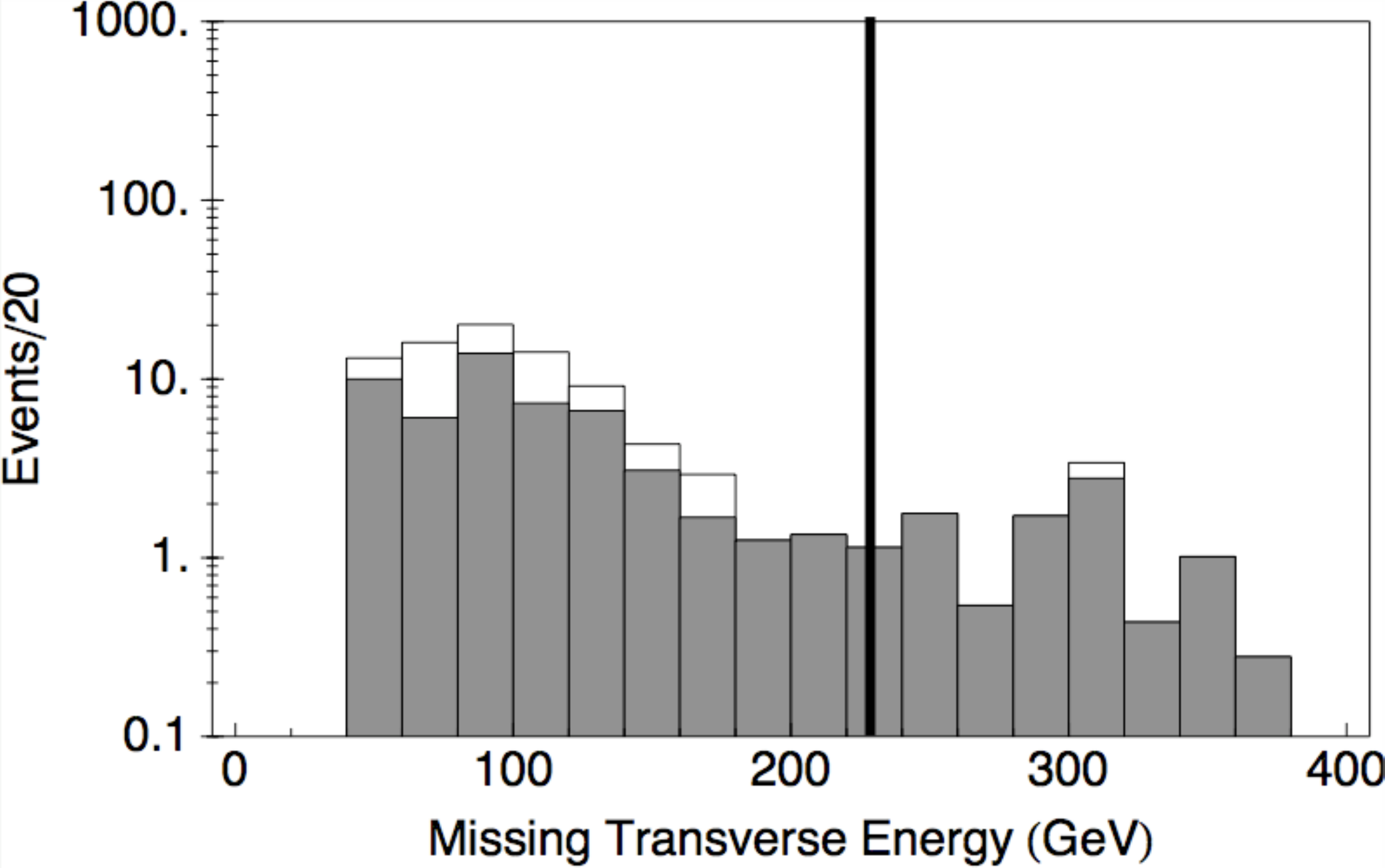}
    \includegraphics[width=3.2in]{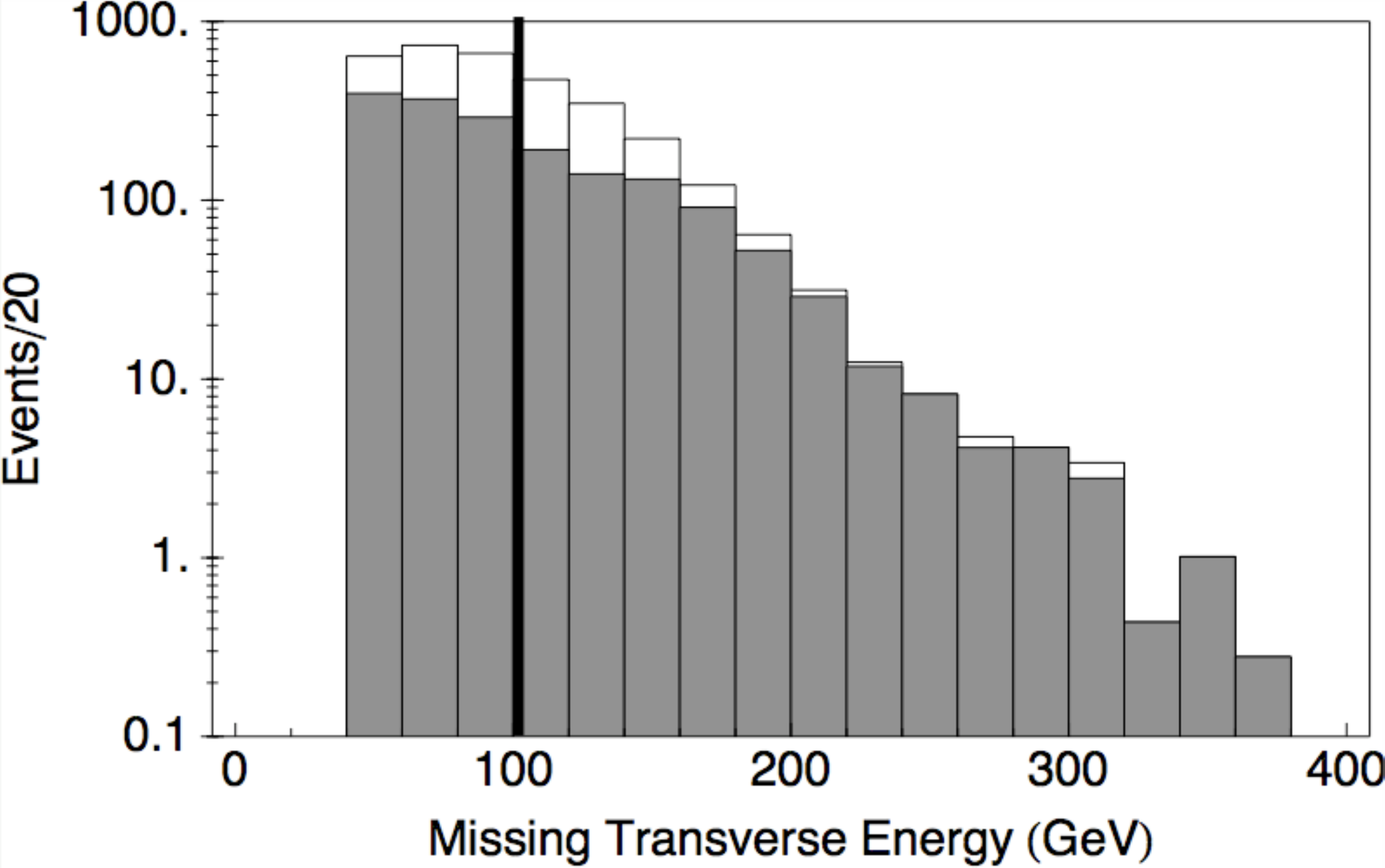}
   \caption{Comparison of \DO cuts and optimized cuts for a sample dijet signal for $m_{\tilde{g}} = 210 \GeV$ and $m_{\widetilde{B}} = 100 \GeV$.  Background distribution is shown in gray and signal distribution in white.  (Left) Using the \DO cuts $H_T \geq 300 \GeV$ and $\MET \geq 225 \GeV$ (Right) Using the more optimal cuts $H_T \geq 150 \GeV$ and $\MET \geq 100 \GeV$.  The optimized cuts allow us to probe regions with larger $S/B$.}
   \label{Fig: optimizedcuts}
\end{figure}

This is explicitly illustrated in Fig. \ref{Fig: optimizedcuts}, which shows the $\MET$ distribution of a dijet sample passed through two different sets of $\MET$ and $H_T$ cuts.  The signal, a 210$\GeV$ gluino directly decaying (i.e., no cascade) to a 100$\GeV$ bino, is shown in white and the Standard Model background, in gray.  The plot on the left shows the events that survive a $300 \GeV$ $H_T$ cut.  While the $H_T$ cut significantly reduces the background, it also destroys the signal above the $\MET$ cut of 225$\GeV$.  These cuts were used in the $\DO$ dijet search;  they are optimized for a $\sim  400\GeV$ gluino, but are clearly not ideal for the signal point shown here.
A more optimal choice of cuts is shown on the right.  While the lower $H_T$ cut of $150 \GeV$ keeps more background, it also keeps enough signal for a reasonable $S/B$ ratio at low $\MET$.  Therefore, with a $\MET$ cut of $100 \GeV$, exclusion limits on this point in parameter space can be placed.     


A model-independent search should have broad acceptances over a wide range of kinematical parameter space.  Ideally, searches should be sensitive to all possible kinematics by considering all appropriate $\MET$ and $H_T$ cuts.  
This can be effectively done by plotting the differential cross section as a function of $\MET$ and $H_T$,
\begin{equation}
\frac{d^2 \sigma}{d H_T d \MET} \Delta H_T \Delta \MET.
\end{equation}
In this case, the results of a search would be summarized in a grid, where each box contains the measured cross section within a particular interval of $\MET$ and $H_T$.

As an example, the differential cross section grids for exclusive $1j - 4^+j$ searches (see Table \ref{Fig: cutstable} for jet selection criteria) at the Tevatron are shown in Table \ref{Fig: diffxsectionlimits}.  The grids are made for the Standard Model background, which include $W^{\pm}+ nj$, $Z^0 + nj$, and $t \bar{t} + n j$. 
The QCD background was not simulated; we expect the QCD contributions to be important for points in the lowest $\MET$ bin.  For details concerning the Monte Carlo generation of the backgrounds, see Sec. \ref{Sec: Background}. 

From these results, it is straightforward to obtain limits on the differential cross section for any new physics signal.  Consider a specific differential cross section measurement that measures $N_m$ events in an experiment.  The Standard Model predicts $B$ events, while some specific theory predicts $B+S$ events, where $S$ is the number of signal events.  
 \begin{table}[t]
\begin{center}
\begin{tabular}{|c||c|c|c|c|}
\hline
&$1j+\SmMET$&$2j+\SmMET$&$3j+\SmMET$&$4^+j+\SmMET$\\
\hline\hline
$E_{T\,j_1}$&$\ge 150$&$\ge 35$&$\ge 35$&$\ge 35$\\
$E_{T\,j_2}$&$< 35$&$\ge 35$&$\ge 35$&$\ge 35$\\
$E_{T\,j_3}$&$< 35$&$<35$&$\ge 35$&$\ge 35$\\
$E_{T\,j_4}$&$< 20$&$<20$&$<20$&$\ge 20$\\
\hline
\end{tabular}
\caption{\label{Tab: Cuts} Summary of the selection criteria for
the four exclusive (i.e., non-overlapping) searches.  The two hardest jets are required to be central ($| \eta | \leq 0.8$).  All other jets must have $| \eta | \leq 2.5$.}
\label{Fig: cutstable}
\end{center}
\end{table}
\begin{table}[h!] 
   \centering
   \begin{tabular}{|c||c|c|}
   \hline
   &Simulated Background& Signal Limits\\
   \hline
   \hline
   ~~\rotatebox{90}{\hspace{0.165in}Monojet}~~~~&
   \includegraphics[width=3in]{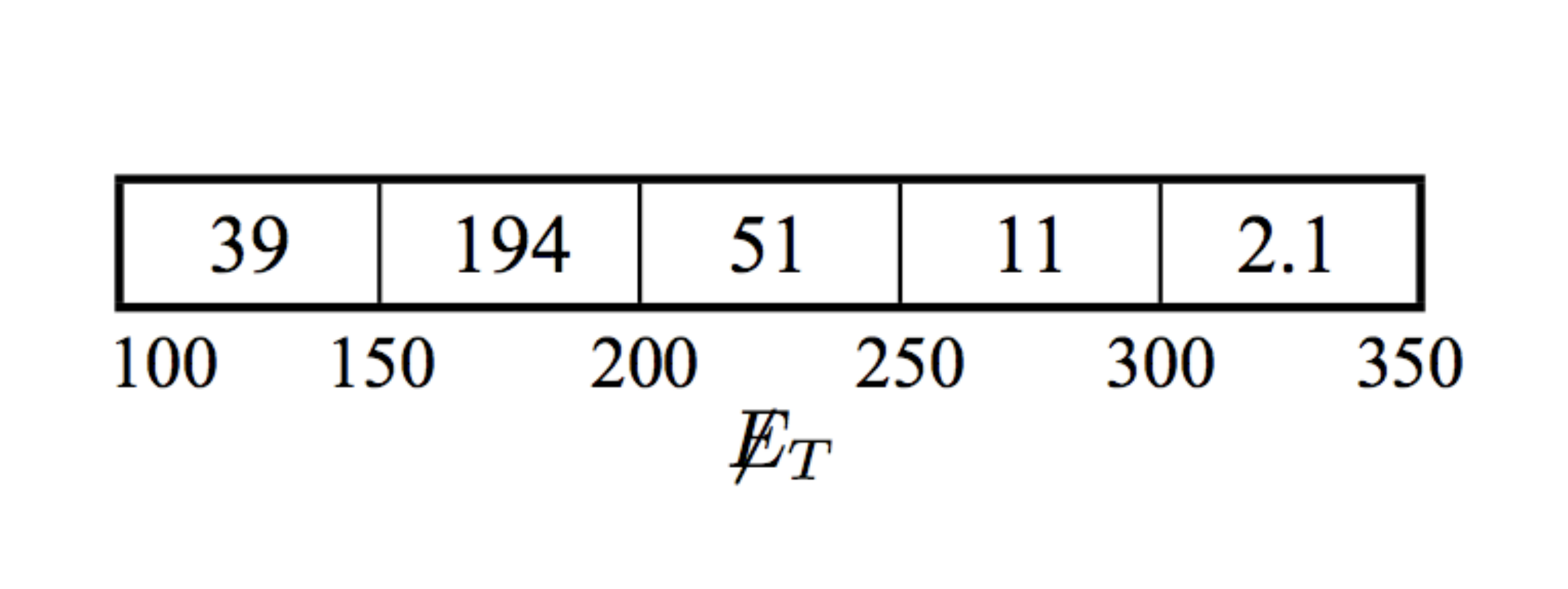}  &
   \includegraphics[width=3in]{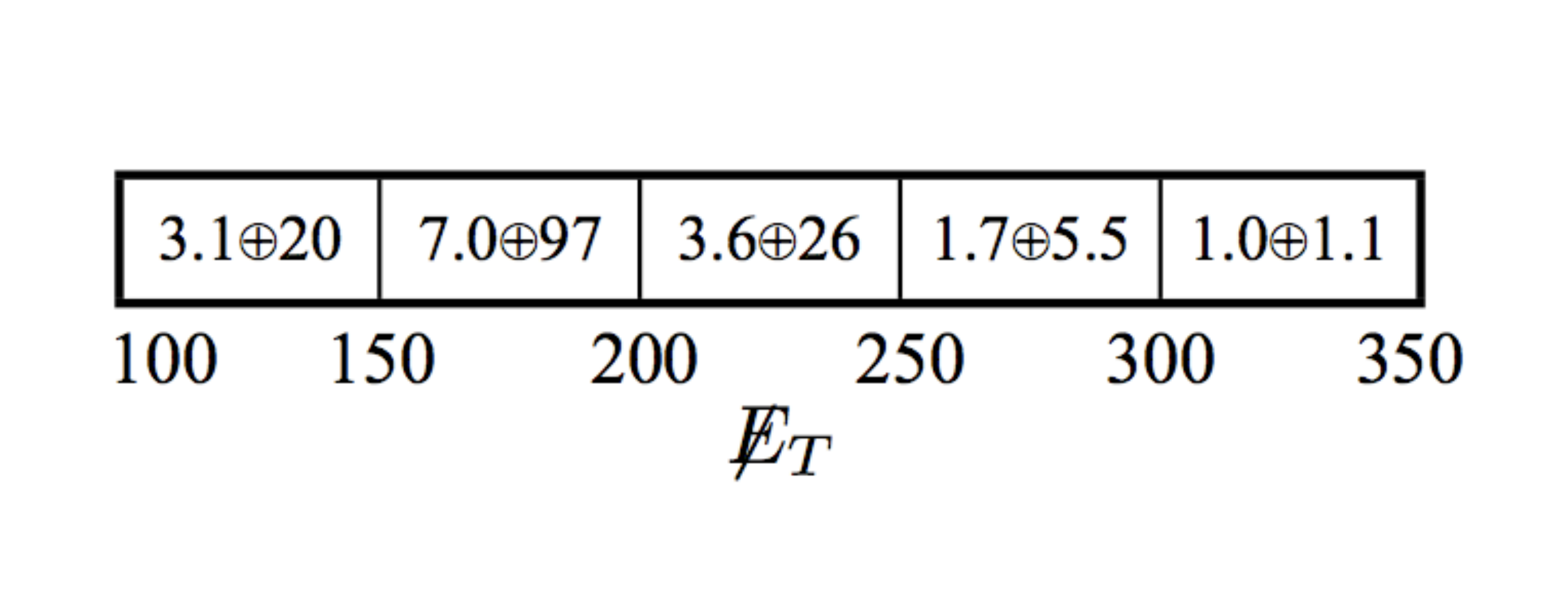}\\
    \hline
   ~~\rotatebox{90}{\hspace{0.57in}Dijet}~~~~&
   \includegraphics[width=3in]{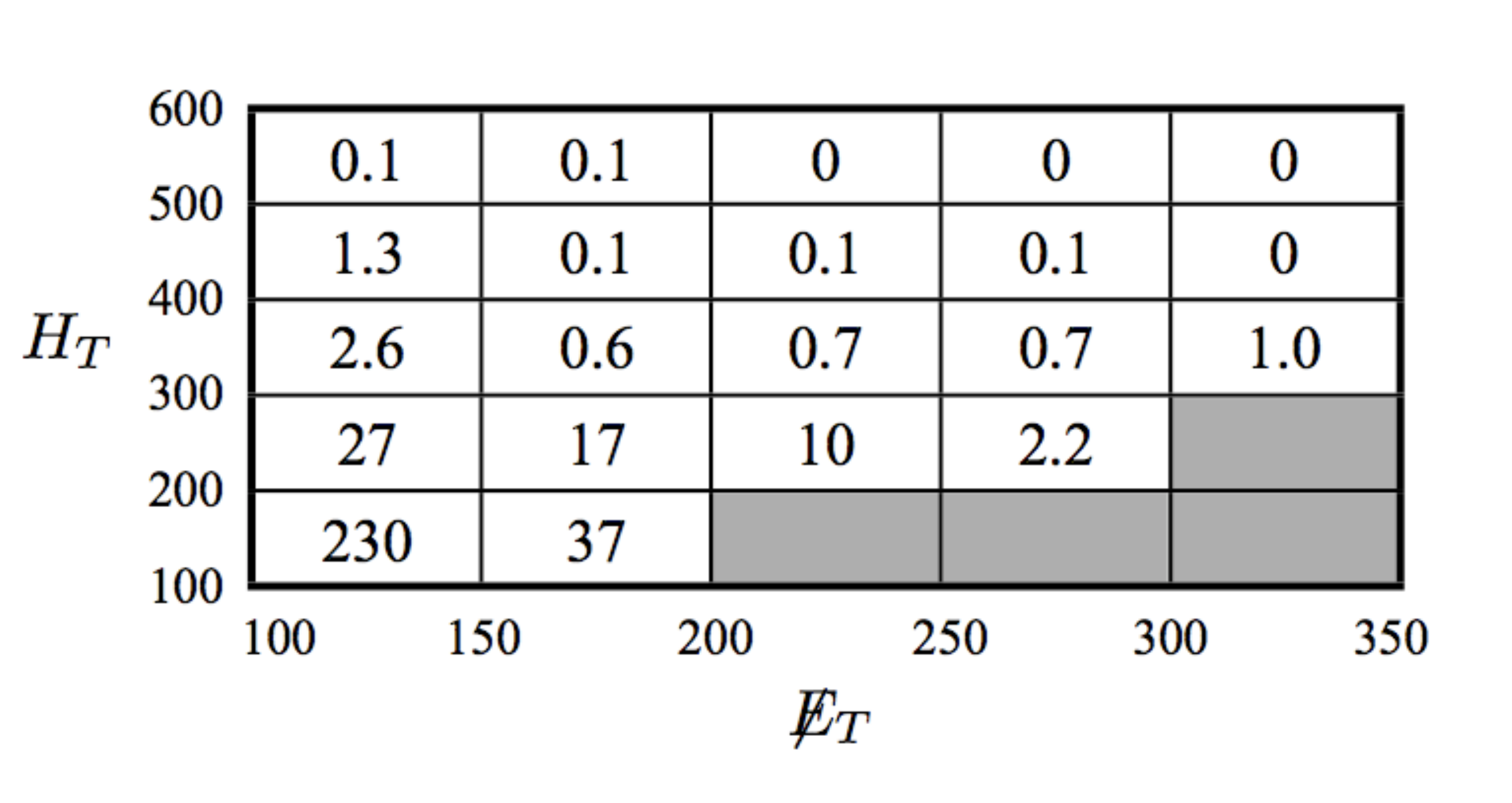}  &
    \includegraphics[width=3in]{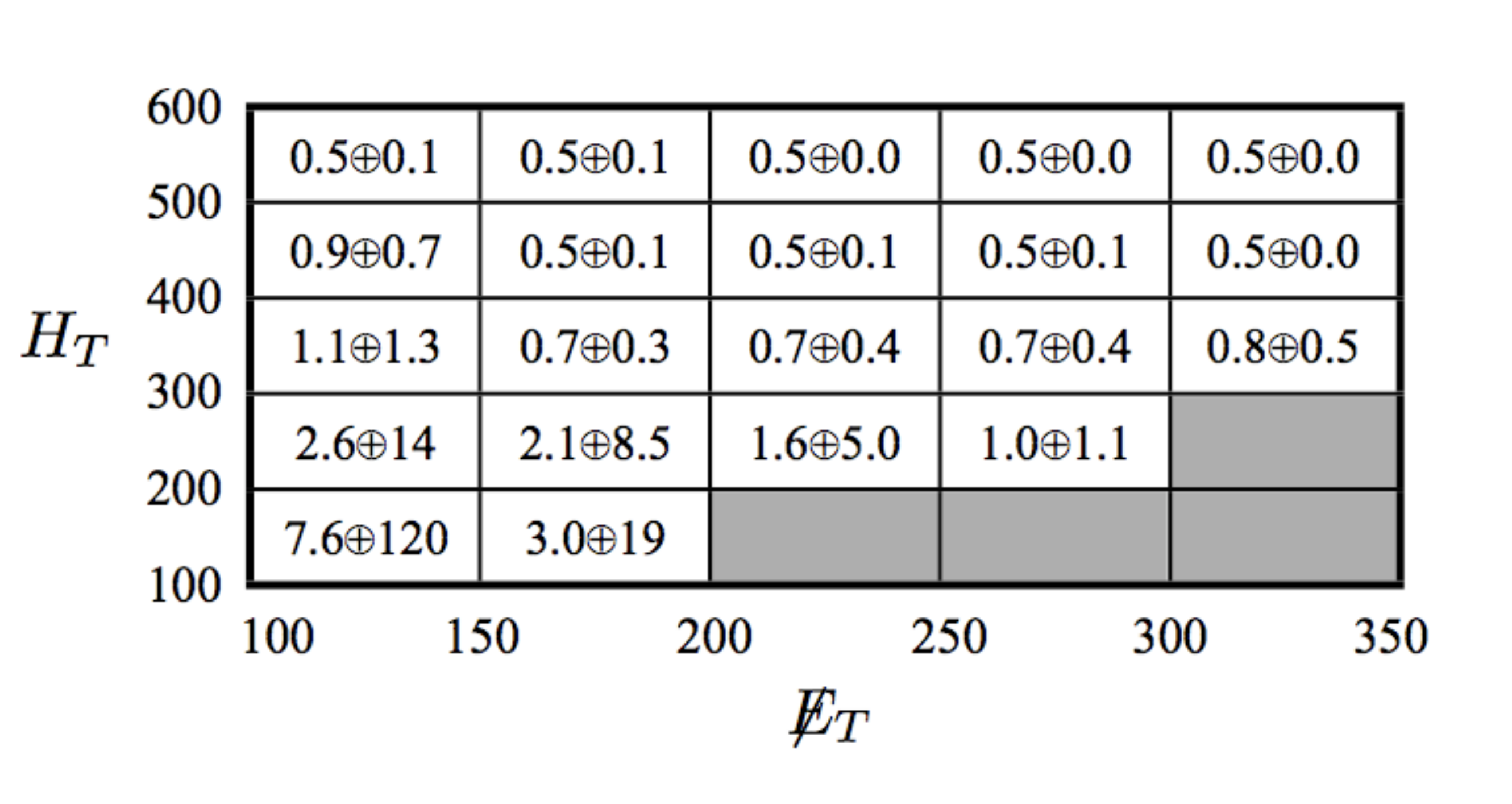}\\
    \hline
 \rotatebox{90}{\hspace{0.4in}Threejet}&
     \includegraphics[width=3in]{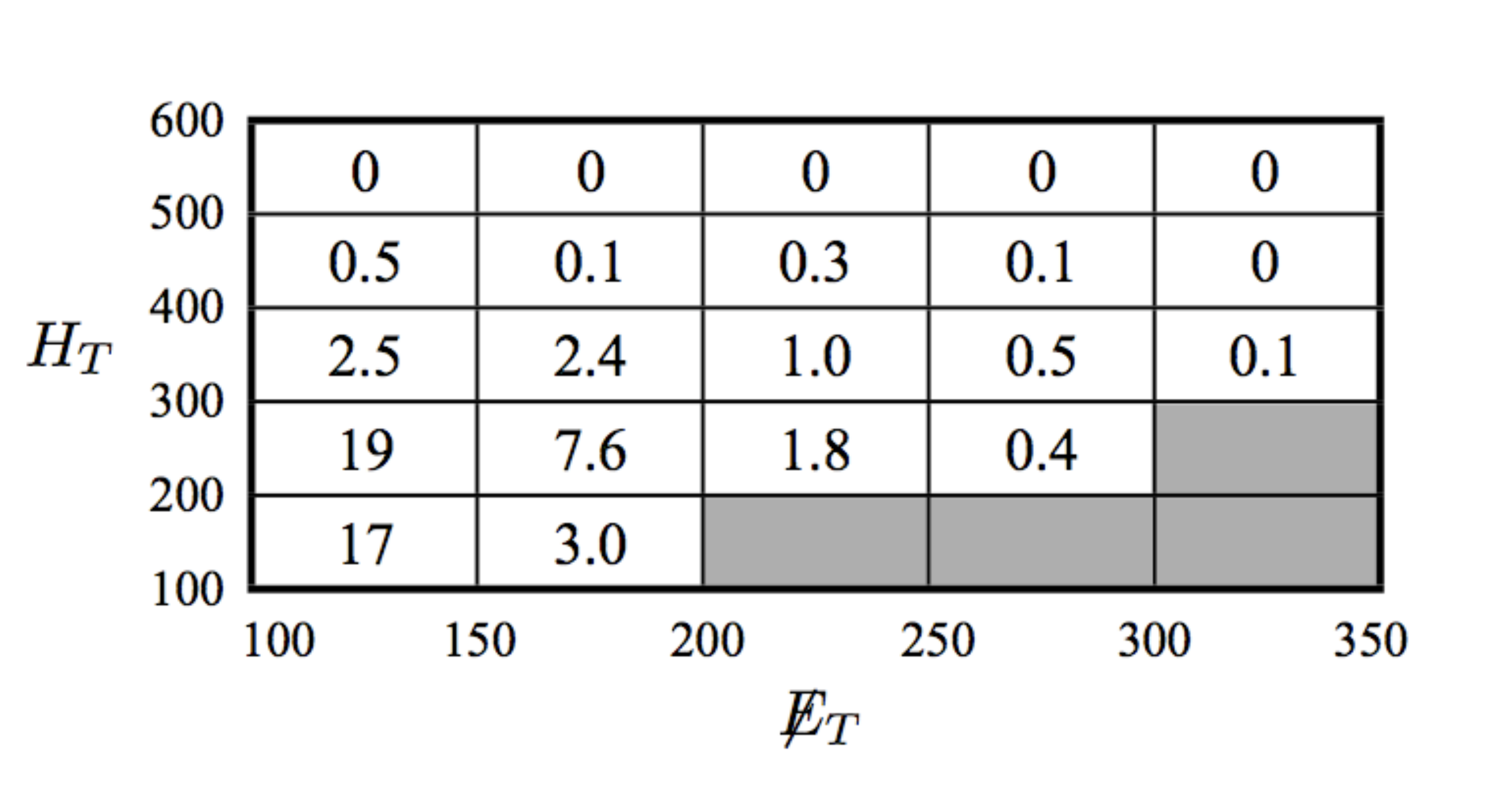}  &
     \includegraphics[width=3in]{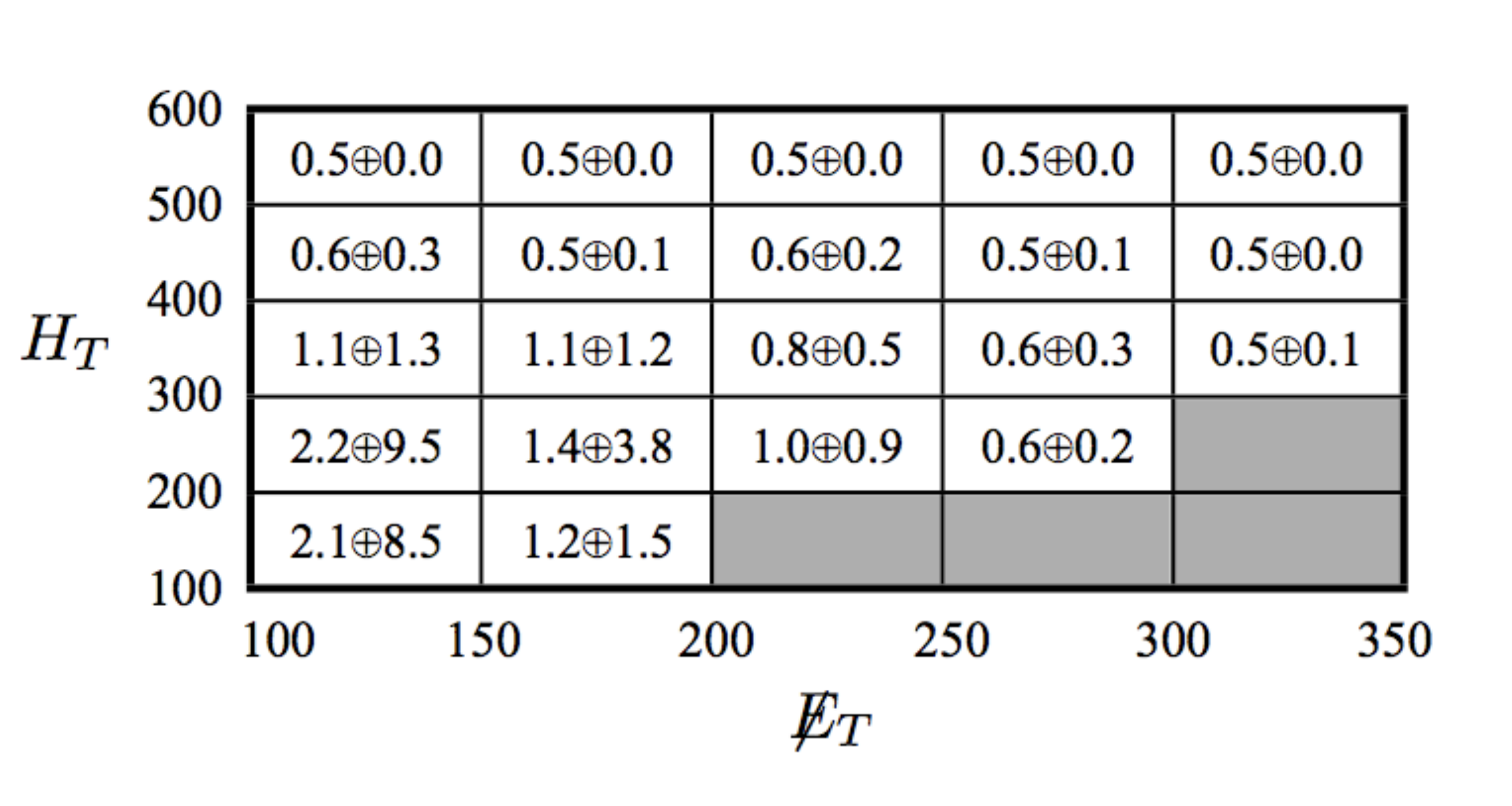}\\
     \hline
 \rotatebox{90}{\hspace{0.45in}Multijet}&
       \includegraphics[width=3in]{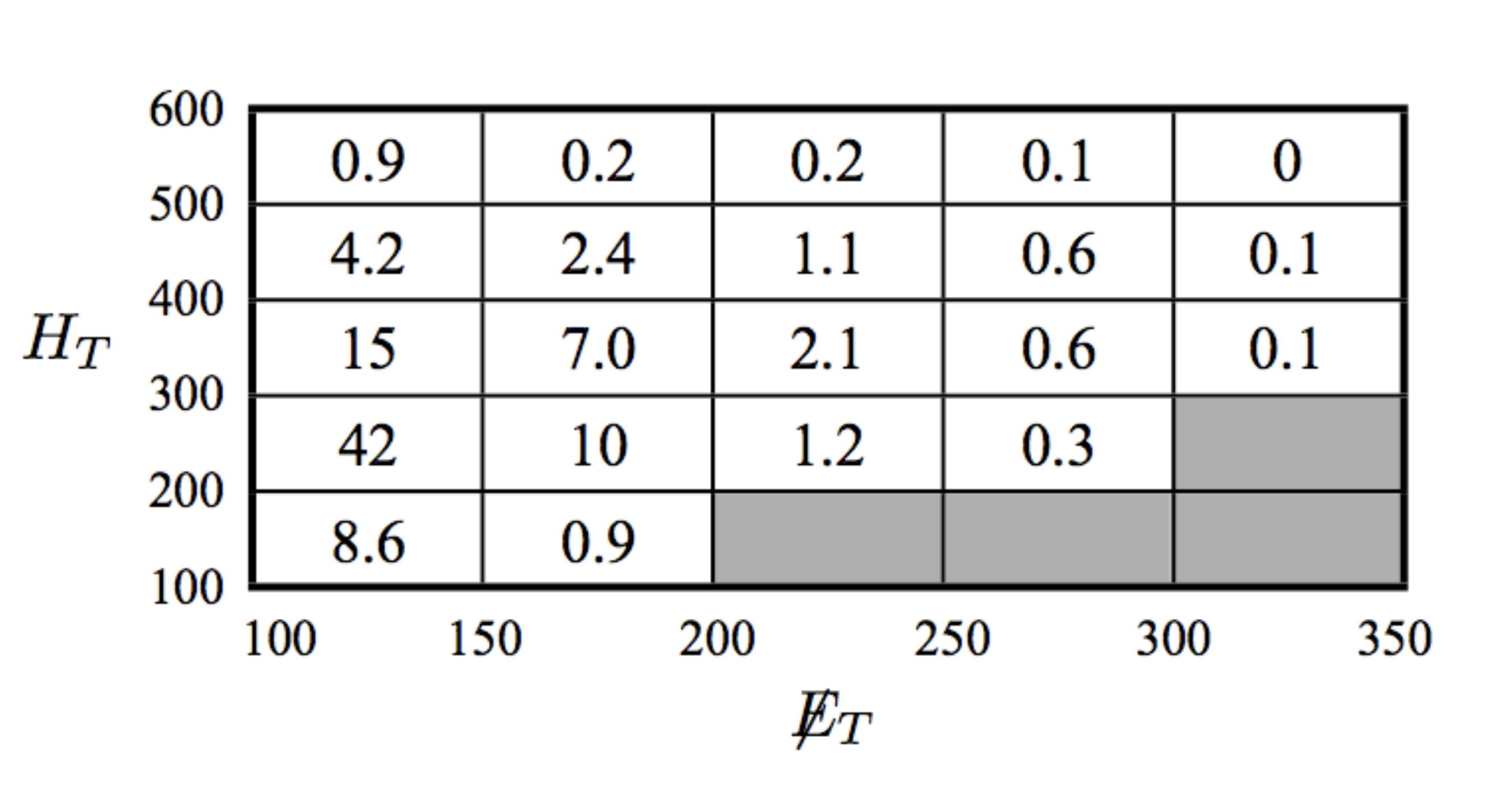}&
         \includegraphics[width=3in]{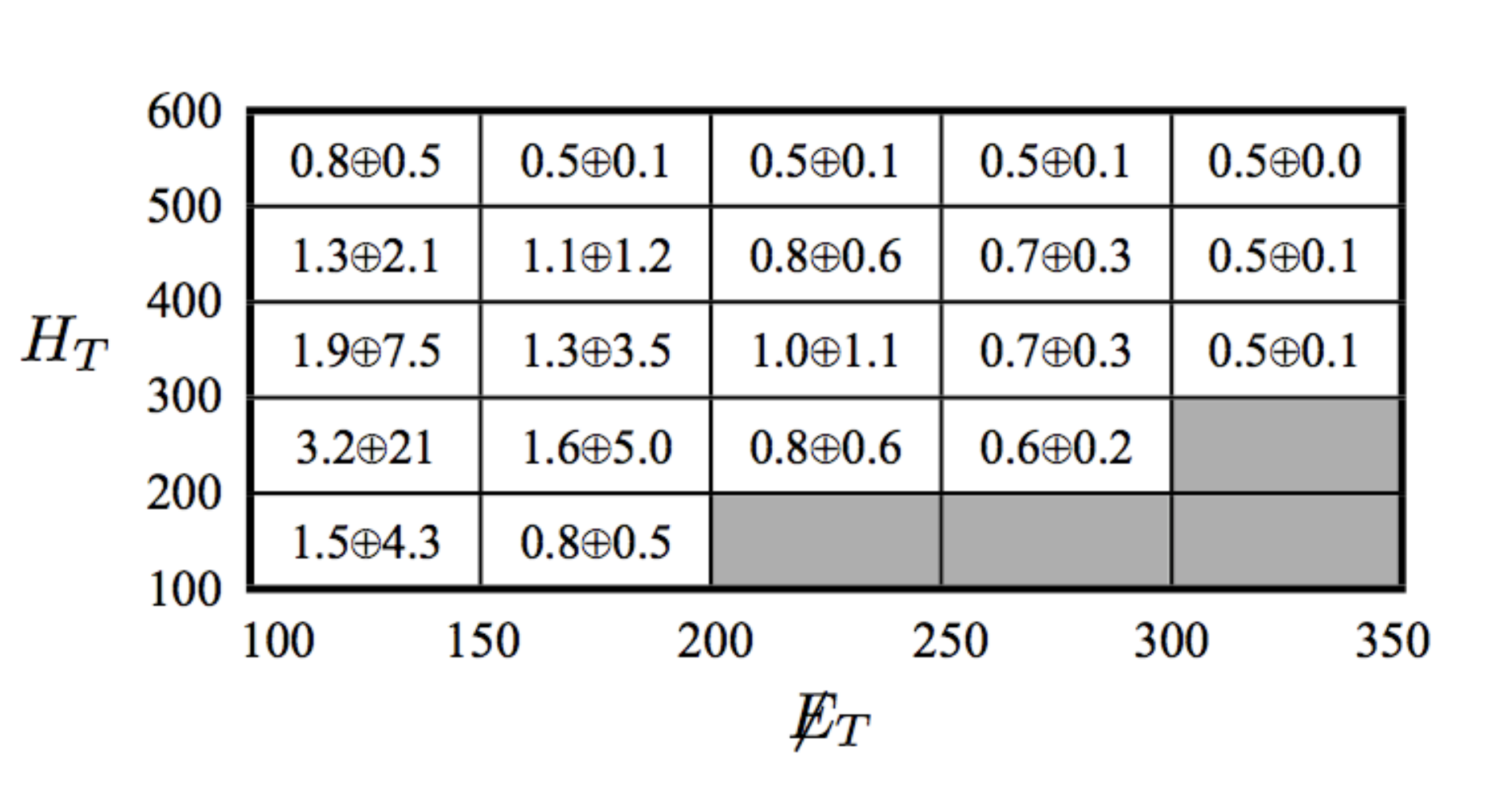}\\
  \hline
\end{tabular}
   \caption{Differential cross section (in fb) for the Standard Model background is shown in the left column for exclusive $1j - 4^+j$ searches.  The expected signal sensitivity at 84\% confidence is shown on the right (in fb).  The statistical error is shown to the left of the $\oplus$ and the systematic error is on the right.  For purposes of illustration, we assume a 50\% systematic error on the background.  The gray boxes are kinematically forbidden.  These results are for 4 fb$^{-1}$ luminosity at the Tevatron.}
   \label{Fig: diffxsectionlimits}
\end{table}
The probability of measuring $n$ events is given by the Poisson distribution 
with mean $\mu = B + S$.  The mean $\mu$ is excluded to 84$\%$ such that
\begin{equation}
e^{-\mu^{\text{excl}}} \sum_{n=0}^{N_m} \frac{(\mu^{\text{excl}})^n}{n!}  \leq 0.16.
\end{equation}
The solution to this equation gives the excluded number of signal events
\begin{equation}
S^{\text{excl}}(N_m, B) = \mu^{\text{excl}}(N_m) - B. 
\end{equation}
The expected limit on the signal is then given by
\begin{equation}
\langle S^{\text{excl}}(B) \rangle = \sum_{N_m=0}^{\infty} S^{\text{excl}}(N_m, B) \frac{e^{-B} B^{N_m}}{N_m!}.
\end{equation}
In the limit of large $B$, the probability distribution approaches a Gaussian and we expect that
\begin{equation}
\lim_{B \rightarrow \infty} \langle S^{\text{excl}}(B) \rangle = \sqrt{B}.
\end{equation}
In the limit of small $B$, we expect that
\begin{equation}
\lim_{B \rightarrow 0} \langle S^{\text{excl}}(B) \rangle = -\ln(0.16) \approx 1.8.
\end{equation}
The right column of Table \ref{Fig: diffxsectionlimits} shows the limit on the differential cross section for any new physics process.   When presented in this fashion, the experimental limits are model-independent and versatile.  
With these limits on the differential cross section, anyone can compute the cross section for a specific model and make exclusion plots using just the signal limits shown in Table \ref{Fig: diffxsectionlimits}.  For the comparison to be reliable, the detector simulator should be properly calibrated.  

In addition to the statistical uncertainty, systematic uncertainties can also be important.
Unlike the statistical uncertainties, the systematic uncertainties can be correlated with each other.
One important theoretical uncertainty is the higher-order QCD correction to the backgrounds.  These QCD uncertainties result in K-factors that change the normalization of the background, but do not significantly alter the background shapes with respect to $H_T$ and $\MET$.  Because this uncertainty is highly correlated between different differential cross section measurements, treating the uncertainty as uncorrelated reduces the 
sensitivity of the searches.  If a signal changes the shape of the differential cross section, e.g. causing a peak in the distribution,  higher order corrections would be unlikely to
explain it.  To make full use of the independent differential cross section measurements, a complete error
correlation matrix should be used.  In practice, because the backgrounds are steeply falling with respect to $H_T$ and $\MET$, assigning an uncorrelated
systematic uncertainty does not significantly hurt the resolving power of the experiment.
In Table \ref{Fig: diffxsectionlimits}, we have assigned a systematic uncertainty of $\epsilon_{\text{sys}} = 50\%$ to
each measurement, which should be added in quadrature to the statistical uncertainty.   This roughly corresponds to the requirement that the total signal to background ratio is one. 
 
The reduced chi-squared $\chi^2_N$ value for $N$ measurements is
\begin{equation}
\chi^2_{N} = \sum_{j=1}^{N} \frac{S_j^2}{(\text{SL}_j)^2 + (\epsilon_{\text{sys}} \times B_j)^2} \times \frac{1}{N},
\end{equation}
where $S_j$ is the number of signal events and $B_j$ is the number of background events in the $j^{\text{th}}$ box of the grid.  The statistical error $\text{SL}_j$ and the systematic error $\epsilon_{\text{sys}}\times B_j$ is read off from Table \ref{Fig: diffxsectionlimits}.  In order to have a useful significance limit, it is necessary to only include measurements where there is an expectation of statistical significance; otherwise, the $\chi^2_N$ is diluted by a large number of irrelevant measurements.  There is no canonical way of dealing with this elementary statistical question, although the $\text{CL}_S$ method is the most commonly used \cite{Junk:1999kv,Read:2000ru}.  In this article, we take a very simple approach.  If the expected significance for a single measurement is
greater than a critical number, $S^{\text{crit}}$, it is included in the $\chi^2_N$, otherwise it is not.
We tried several values of $S^{\text{crit}}$ and the experimental sensitivity to different theories was not altered by the different choices.  We chose $S^{\text{crit}}= 0.5$ for the exclusion plots.  This method does not maximize the reach in all cases, but because there are usually just a few measurements that give large significance, we are relatively insensitive to the exact statistical procedure.

In what follows, we will apply the general philosophy presented here to find the exclusion region for gluinos that are pair-produced at the Tevatron.\footnote{Throughout this article, ``gluino" refers to a color octet fermion, ``wino" to a charged SU(2) fermion, and ``bino" to a neutral singlet.  These names imply nothing more than a particle's quantum numbers.}  In Sec. \ref{Sec: Searches}, we will explain how the signal and background events have been generated.  In Sec. \ref{Sec: exclusion}, we will show how  mass bounds can be placed on the gluino and bino masses using the proposed model-independent analysis and will discuss the challenges presented by cascade decays.  We conclude in Sec. \ref{Sec: conclusion}.

\section{Event Generation}
\label{Sec: Searches}

\subsection{Signal}

In this section, we discuss the generation of signal events for the gluino cascade decay shown in (\ref{eq: decay}).  The experimental signatures of this decay chain are determined primarily by the spectrum of particle masses.  In particular, the mass splittings determine how much energy goes into the jets as opposed to the bino - i.e., the ratio of the visible energy to missing transverse energy.  Events with large $H_T$ and $\MET$ will be the easiest to detect; this is expected, for example, when a heavy gluino decays into a wino that is nearly degenerate with either the gluino or the bino. 
 The reach of the searches is degraded, however, when the wino is included as an intermediate state in the decay chain.  When the jets from the cascade decay are all hard, the missing energy is significantly smaller than what it would be for the direct decay case.  Picking out signals with small missing transverse energy is challenging because they push us closer to a region where the dominant background is coming from QCD and is poorly understood.  This happens, in particular, when the mass splitting between the gluino and bino is large and the wino mass is sufficiently separated from both.  When the wino is nearly degenerate with either the gluino or the bino, then we expect to see $2$ hard jets and $2$ soft jets from the decay.  This case begins to resemble the direct decay scenario; there is more missing energy and, therefore, the signal is easier to see.
It is particularly challenging to probe regions of parameter space where the gluino is nearly degenerate with the bino.  For this case, even in the light-gluino region ($m_{\tilde{g}} \lesssim 200 \GeV$), the benefit of the high production cross section for the gluinos is overwhelmed by the  small missing transverse momentum in each event; the jets in these events are soft and the $p_T$ of the two binos approximately cancel when summed together \cite{Alwall:2008ve}.  Even if the gluinos are produced at large invariant mass, the situation is not markedly improved; in this case, the jets from each gluino are collinear and aligned with the $\MET$.    
Such events are easily mistaken as QCD events and eliminated by the cuts that are implemented to reduce the QCD backgrounds.

The inclusion of hard initial-state jets significantly increases the exclusion reach in this degenerate region of parameter space.  The initial-state radiation boosts the gluinos in the same direction, decreasing the angle between them, which in turn, enhances the $\MET$.  Therefore, ISR jets allow us to capitalize on the high production cross section of light gluinos to set bounds on their masses.  

To properly account for initial-state radiation (ISR) and final-state radiation (FSR), MadGraph/MadEvent \cite{Alwall:2007st} was used to generate events of the form
\begin{equation}
p \bar{p} \rightarrow \tilde{g} \tilde{g} + N j,
\end{equation}
where $N=0,1,2$ is the multiplicity of QCD jets.  Pythia 6.4
\cite{Sjostrand:2006za} was used for parton showering and
hadronization.  Properly counting the number of events after parton
showering requires some care.  In general, an $(n+1)$-jet event can be
obtained in two ways: by a $(n+1)$ hard matrix-element, or by hard
radiation emitted from an $n$-parton event during showering.  It is
important to understand which of the two mechanisms generates the
$(n+1)$-jet final state to ensure that events are not double-counted.

In this article, a version of the so-called MLM matching procedure
implemented in MadGraph/MadEvent and Pythia \cite{Alwall:2007fs} was
used for properly merging the different parton multiplicity
samples. This matching has been implemented both for Standard Model
production and for beyond the Standard Model processes. In this
procedure, parton-level events are generated with a matrix element
generator with a minimum distance between partons characterized by the
$k_{\perp}$ jet measure:
\begin{eqnarray} 
\nonumber
&&d^2(i,j) = \Delta R^2_{ij} \min(p_{Ti}^2, p_{Tj}^2)\\
&&d^2(i,\text{beam}) = p_{Ti}^2,
\end{eqnarray}
where $\Delta R^2_{ij} = 2[\cosh(\Delta \eta) - \cos(\Delta \phi)]$ \cite{jetmeasure}.  
The event is clustered using the $k_T$ clustering algorithm,
allowing only for clusterings consistent with diagrams in the matrix
element, which can be done since MadGraph generates all diagrams for
the process. The $d^2$ values for the different clustered vertices are
then used as scales in the $\alpha_s$ value corresponding to that
vertex, i.e.\ the event weight is multiplied by
$\prod_i\frac{\alpha_s(d^2_i)}{\alpha_s(\mu_R^2)}$, where the product
is over the clustered vertices $i$. This is done in order to treat
radiation modeled by the matrix element as similarly as possible to
that modeled by the parton shower, as well as to correctly include a tower of next-to-leading log terms.
A minimum cutoff $d(i,j) > Q_{\text{min}}^{\text{ME}}$ is placed on
all the matrix-element multi-parton events. 

After showering, the partons are clustered into jets using the standard $k_{\perp}$ algorithm.  
Then, the jet closest to the hardest parton in $(\eta,\phi)$-space is
selected.  If the separation between the jet and parton is within some
maximum distance, $d(\text{parton}, \text{jet}) <
Q_{\text{min}}^{\text{PS}}$, the jet is considered matched.  The
process is repeated for all other jets in the event.  In this way,
each jet is matched to the parton it originated from before showering.
If an event contains unmatched jets, it is discarded, unless it is the
highest multiplicity sample.  In this case, events with additional
jets are kept, provided the additional jets are softer than the
softest parton, since there is no higher-multiplicity matrix element
that can produce such events.
\begin{figure}[b] 
   \centering
   \includegraphics[width=4.5in]{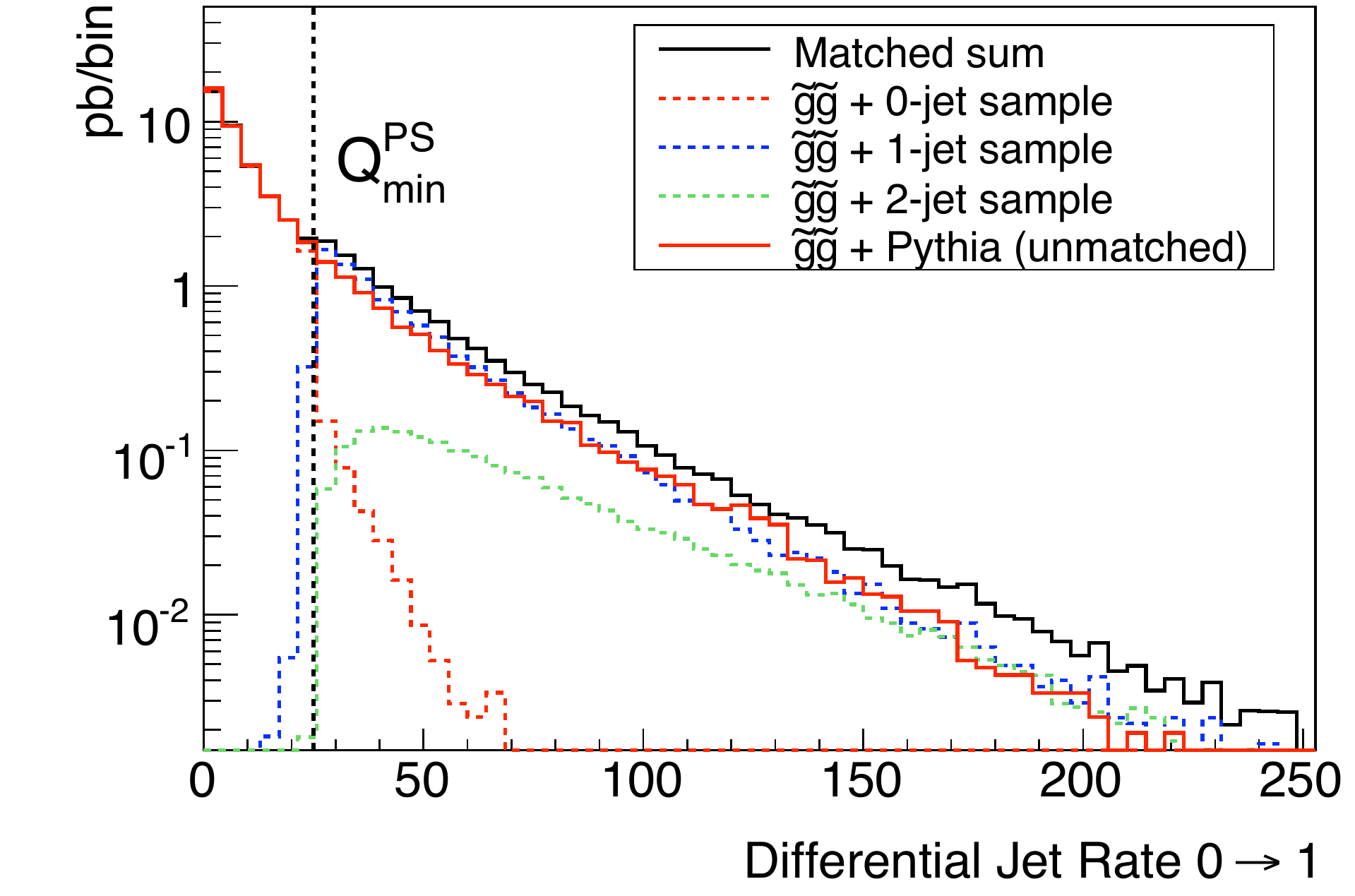}
   \caption{Differential $0\to1$ jet rate for
a matched sample of light gluino production. The full black curve
shows the matched distribution, and the broken curves
show the contributions from different matrix element parton
multiplicity samples. The matching scale $Q_{\text{min}}^{\text{PS}}$
is marked by the dashed line. The full red curve shows the result using
Pythia only.}
\label{Fig: diffjetrate}
\end{figure}
The matching procedure ensures that jets are not double-counted
between different parton multiplicity matrix elements, and should
furthermore give smooth differential distributions for all jet
observables.  The results should not be sensitive to the particular
values of the matching parameters, as long as they are chosen in a
region where the parton shower is a valid description.  Typically, the
matching parameters should be on the order of the jet cuts employed and be far below the factorization scale of the process.  For the gluino production, the parameters were
\begin{equation}
Q_{\text{min}}^{\text{ME}} = 20 \GeV \quad \quad \quad Q_{\text{min}}^{\text{PS}} = 30 \GeV.
\end{equation}
Figure \ref{Fig: diffjetrate} shows the differential jet rate going
from zero to one jets $D(1j \rightarrow 0j)$, which is the maximum
$k_{\perp}$ distance for which a $1j$ event is characterized as a $0j$
event.  Below $Q_{\text{min}}^{\text{PS}}$, all jets come from parton
showering of the $0j$ multiplicity sample.  Above
$Q_{\text{min}}^{\text{PS}}$, the jets come from initial-state
radiation.  The main contributions in this region are from the $1j$
and $2j$ multiplicity samples.  The sum of all the multiplicity
samples is a smooth distribution, eliminating double counting between
the different samples.

The simulations were done using the CTEQ6L1 PDF and with the
renormalization and factorization scales set to the gluino mass
\cite{Pumplin:2002vw}.  The matched cross-sections were rescaled to
the next-to-leading-order (NLO) cross sections obtained using Prospino
2.0.  PGS was used for detector simulation \cite{PGS}, with jets being
clustered according to the cone algorithm, with $\Delta R = 0.5$.  
As a check on this procedure, we compared our results to the signal point given in \cite{Abazov:2007ww} and found that they agreed to within 10$\%$.

\begin{figure}[b] 
   \centering
 \begin{tabular}{|c||cc|c|}
      \hline
   &Matched& &Unmatched\\
   \hline
     \hline
     ~~ &  &  \quad & \\
   ~~\rotatebox{90}{\hspace{0.6in} 40$\GeV$ Bino}~~~~&
   \includegraphics[width=3in]{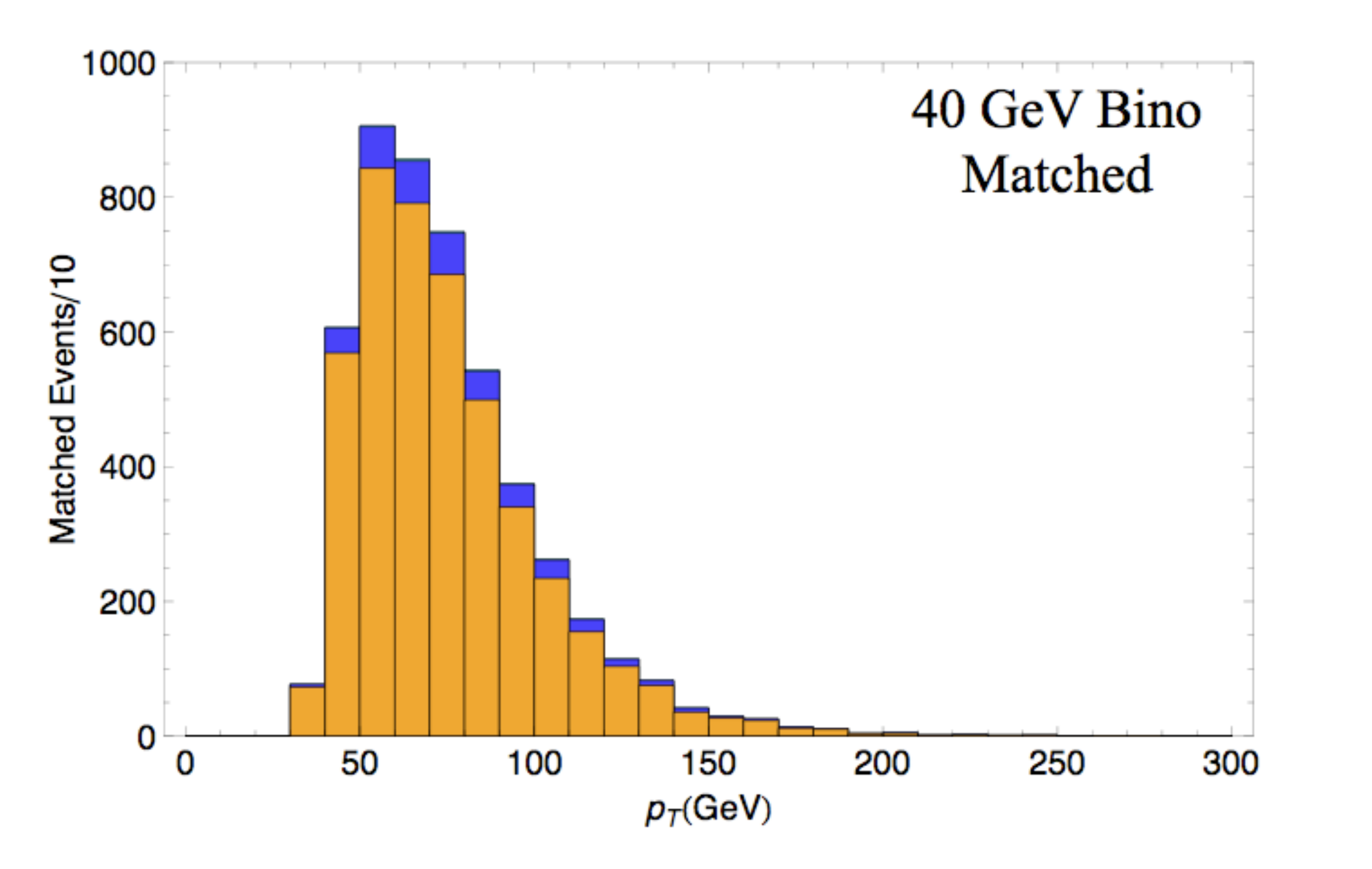} &
     \quad \quad&
         \includegraphics[width=3in]{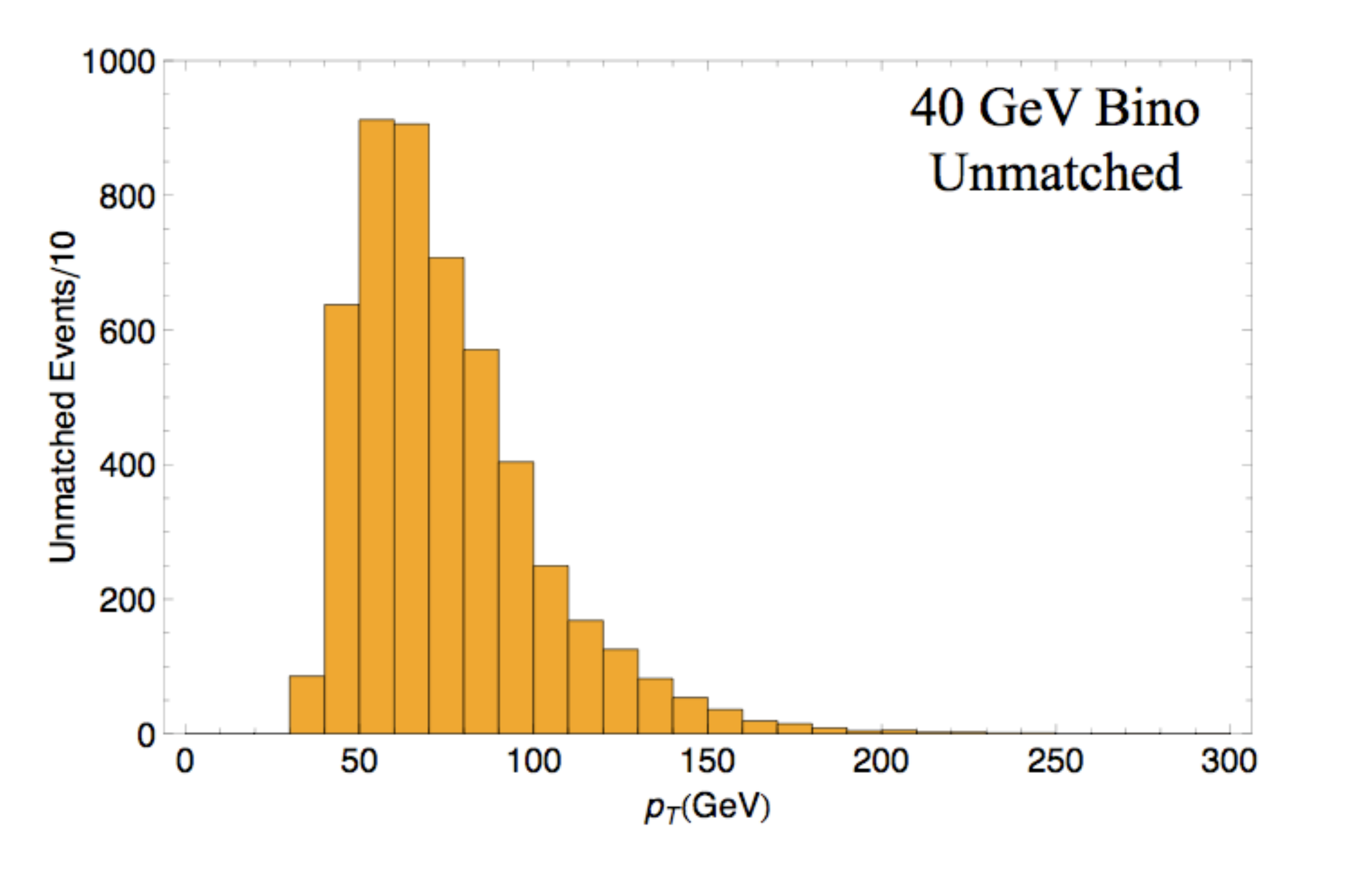} \\
         \hline
              ~~ &  &  \quad & \\
            ~~\rotatebox{90}{\hspace{0.6in}130$\GeV$ Bino}~~~~&
 \includegraphics[width=3in]{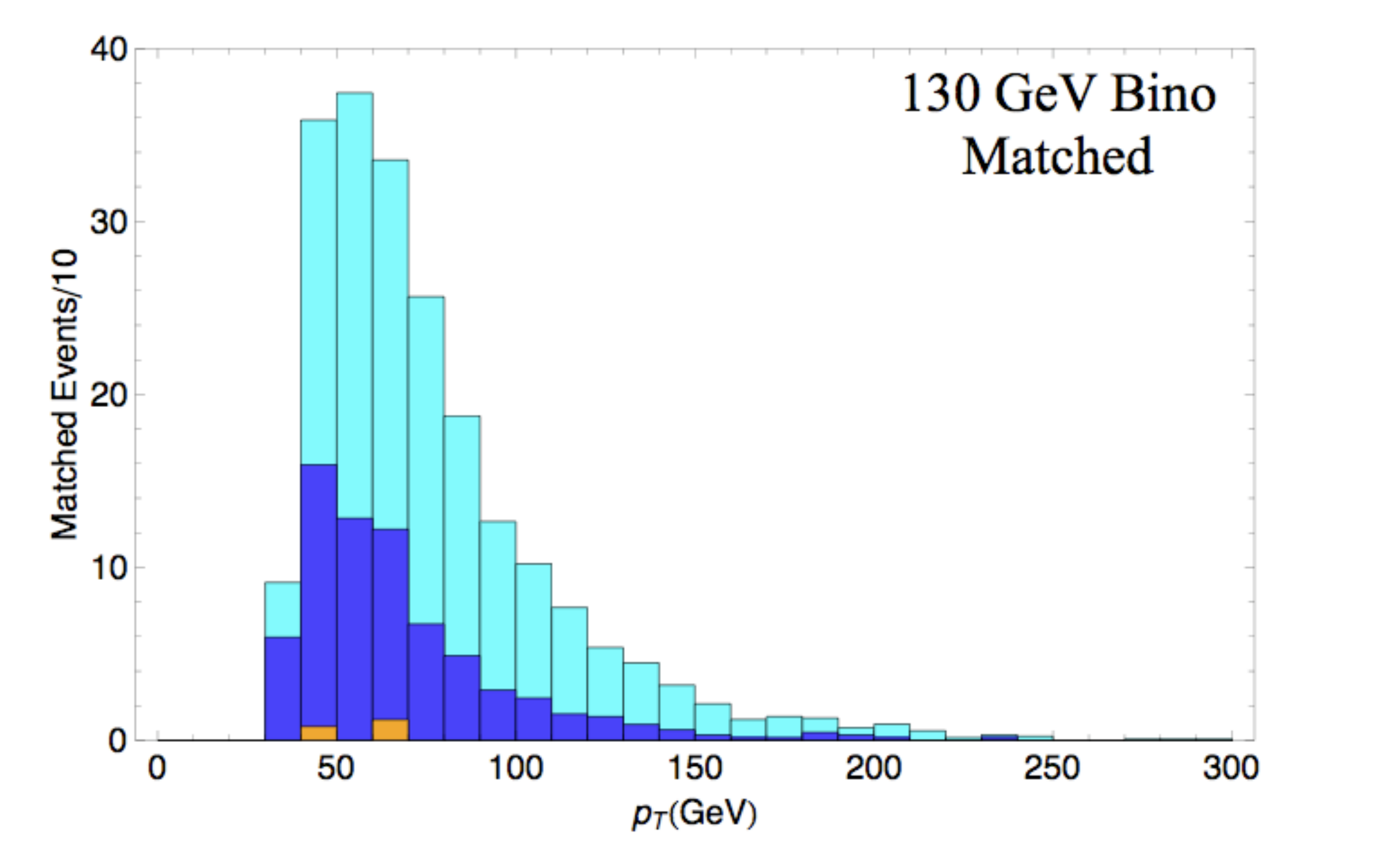} & \quad \quad \quad &
    \includegraphics[width=3in]{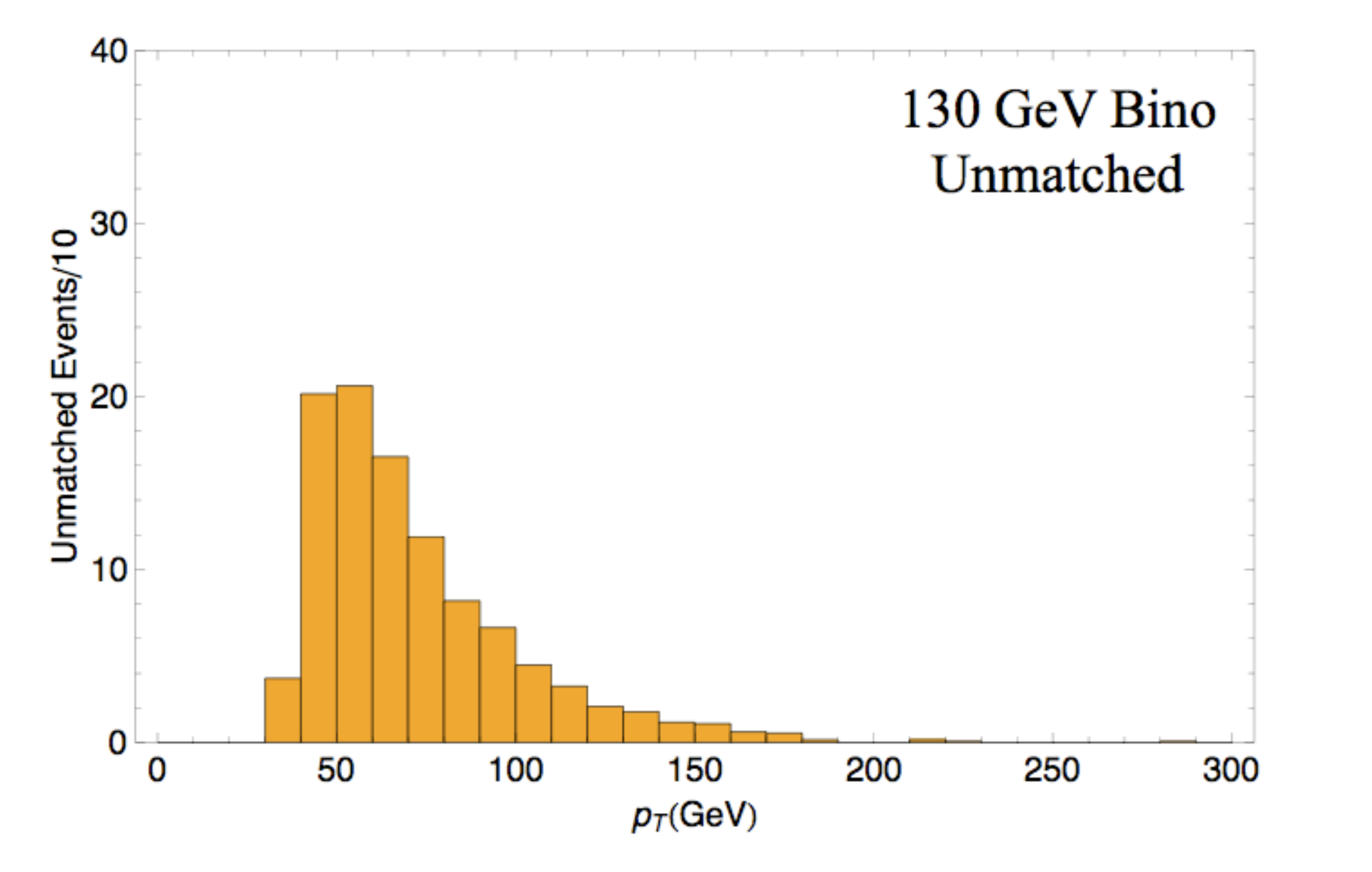}\\
    \hline
\end{tabular}
   \caption{Comparison of matched and unmatched events for a dijet sample of 150$\GeV$ gluinos directly decaying into 40$\GeV$ (top) and 130$\GeV$ (bottom) binos.  The $p_T$ of the hardest jet is plotted in the histograms (1 fb$^{-1}$ luminosity).  Matching is very important in the degenerate case when the contribution from initial state radiation is critical.  The different colors indicate the contributions from 0j (orange), 1j (blue), and 2j (cyan).}
   \label{Fig: pythiamatching}
\end{figure}

To emphasize the importance of properly accounting for initial-state radiation using matching, Fig. \ref{Fig: pythiamatching} compares the $p_T$ distribution for the hardest jet in a matched (left) and unmatched (right)  dijet sample for a 150$\GeV$ gluino directly decaying to a bino.  The colors indicate the contributions from the different multiplicity samples: $0j$ (orange), $1j$ (blue), and $2j$ (cyan).  When the gluino-bino mass splitting is large enough to produce hard jets (top row), the $0j$ multiplicity sample is the main contributor.  ISR is not important in this case and there is little difference between the matched and unmatched plots.  The bottom row shows the results for a 130$\GeV$ bino that is nearly degenerate with the gluino.  In this case, only soft jets are produced in the decay and hard ISR jets are critical for having events pass the dijet cuts.  Indeed, we see the dominance of the $2j$ multiplicity sample in the histogram of matched events.  When ISR is important, the unmatched sample is clearly inadequate, with nearly 60\% fewer events than the matched sample.
  
  \subsection{Backgrounds}
\label{Sec: Background}

The dominant backgrounds for jets + $\MET$ searches are $W^{\pm}/Z^0$ + jets, $t\bar{t}$, and QCD.  Additional background contributions come from single top and di-boson production (WW, WZ, ZZ), but these contributions are sub-dominant, so we do not consider them here.  The missing transverse energy comes from $Z^0 \rightarrow \nu \nu$ and $W^{\pm} \rightarrow l^{\pm} \nu$, where the $W^{\pm}$ boson is produced directly or from the top quark.  To reduce the $W^{\pm}$  background, a veto was placed on isolated leptons with $p_T \geq 10 \GeV$.  However, these cuts do not completely eliminate the $W^{\pm}$ background because it is possible to miss either the electron or muon (or misidentify them).  It should be noted that muon isolation cuts were not placed by PGS, but were applied by our analysis software.  If the muon failed the isolation cut, then it was removed from the record and its four-momentum was added to that of the nearest jet.   Additionally, the $W^{\pm}$ can decay into a hadronic $\tau$, which is identified as a jet.  Because the $\DO$ analysis did not veto on hadronic taus, we have treated all taus as jets in this study.  

QCD backgrounds can provide a significant source of low missing energy events, but are challenging to simulate.  The backgrounds can arise from jet energy mismeasurement due to poorly instrumented regions of the detector (i.e., dead/hot calorimeter cells, jet punch-through, etc.).  Additionally, there are many theoretical uncertainties - for example, in the PDFs, matrix elements, renormalisation, and factorisation/matching scales - that factor in the Monte Carlo simulations of the backgrounds.  For heavy-flavor jets, there is the additional $\MET$ contribution coming from leptonic decays of the b-quarks.  It is possible, for instance, to have the b-quark decay into a lepton and a neutrino, with the neutrino taking away a good portion of the b-quark's energy.  Simulation of the QCD background is beyond the scope of Pythia and PGS and was not attempted in this work.  To account for the QCD background, we imposed a tight lower bound on the $\MET$ of $100 \GeV$.  Jet energy mismeasurement was accounted for by placing a lower bound of $90^{\circ}$ and $50^{\circ}$ on the azimuthal angle between the $\MET$ and the first and second hardest jets, respectively.  In addition, an acoplanarity cut of $165^{\circ}$ was placed between the two hardest jets.  For the dijet case, the azimuthal angle between the $\MET$ and any jet with $p_T \geq 15 \GeV$ and $|\eta| \leq 2.5$ was bounded from below by $40^{\circ}$.  This cut was not placed on the threejet or multijet searches because of the greater jet multiplicity in these cases.

The $W^{\pm}/Z^0 +nj$ and $t \bar{t}$ backgrounds were generated using
MadGraph/MadEvent, with showering and hadronization in PYTHIA.  PGS
was again used as the detector simulator for jet clustering.  The
$W^{\pm}/Z^0$ backgrounds were matched up to 3 jets using the MLM
matching procedure discussed in the previous section, with matching
parameters $Q_{\text{min}}^{\text{ME}} = 10 \GeV$ and
$Q_{\text{min}}^{\text{ME}}=15 \GeV$.  The $t \bar{t}$ backgrounds
were matched up to 2 jets with parameters $Q_{\text{min}}^{\text{ME}}
= 14 \GeV$ and $Q_{\text{min}}^{\text{ME}}=20 \GeV$.  For each of the
separate backgrounds, 500K events were generated.  The results
approximately reproduce the shape and scale of the $\MET$ and $H_T$
distributions published by the $\DO$ collaboration for 1 fb$^{-1}$ \cite{Abazov:2007ww}.
In the dijet case, our results correspond to those of $\DO$ within
$\pm 20\%$.  The correspondence is similar for the $t\bar{t}$
backgrounds in the threejet and multijet cases.  For the $W^{\pm}/Z^0$
backgrounds, the correspondence is within $\pm30 - 40 \%$.  It is
possible that this discrepancy is due to difficulties to fully
populate the tails of the $\MET$ and $H_T$ distributions with good statistics.  In the case of the $W^{\pm}$ background, the modeling of the lepton detection efficiency in PGS might also play a role.  Heavy flavor jet contributions were found to contribute $2\%$ to the $W^{\pm}/Z^0$ backgrounds, which is well below the uncertainties that arise from not having NLO calculations for these processes and from using PGS.

\section{Gluino Exclusion Limits}
\label{Sec: exclusion}
\begin{table}[b] 
   \centering
   \begin{tabular}{|c|c||c|c|}
   \hline
 \multicolumn{4}{|c|}{Sample Model} \\
   \hline
   \hline
   ~~\rotatebox{90}{\hspace{0.38in}Monojet}~~~~ &
   \includegraphics[width=2.5in]{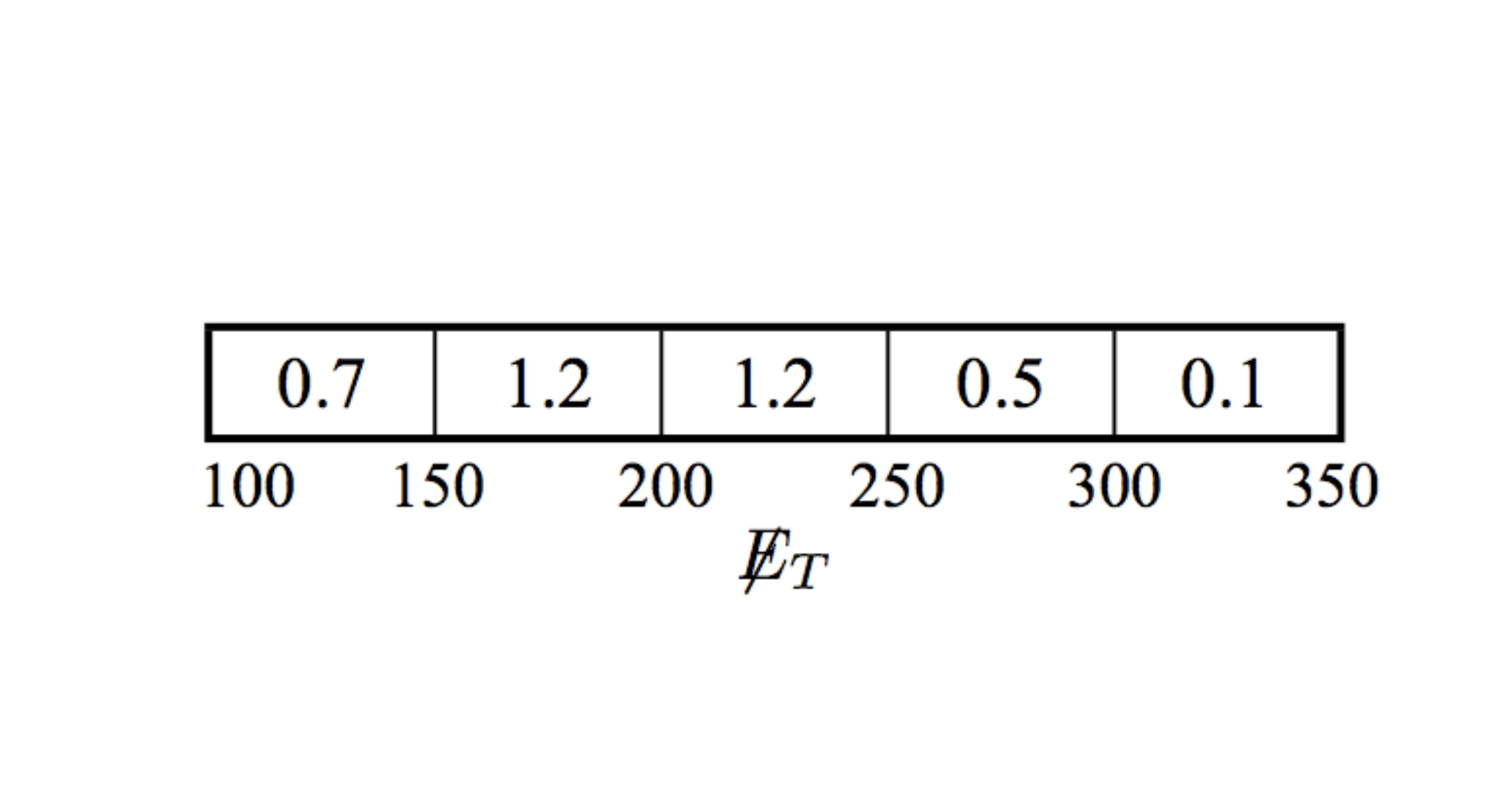}&
   ~~\rotatebox{90}{\hspace{0.47in}Dijet}~~~~ &
    \includegraphics[width=2.5in]{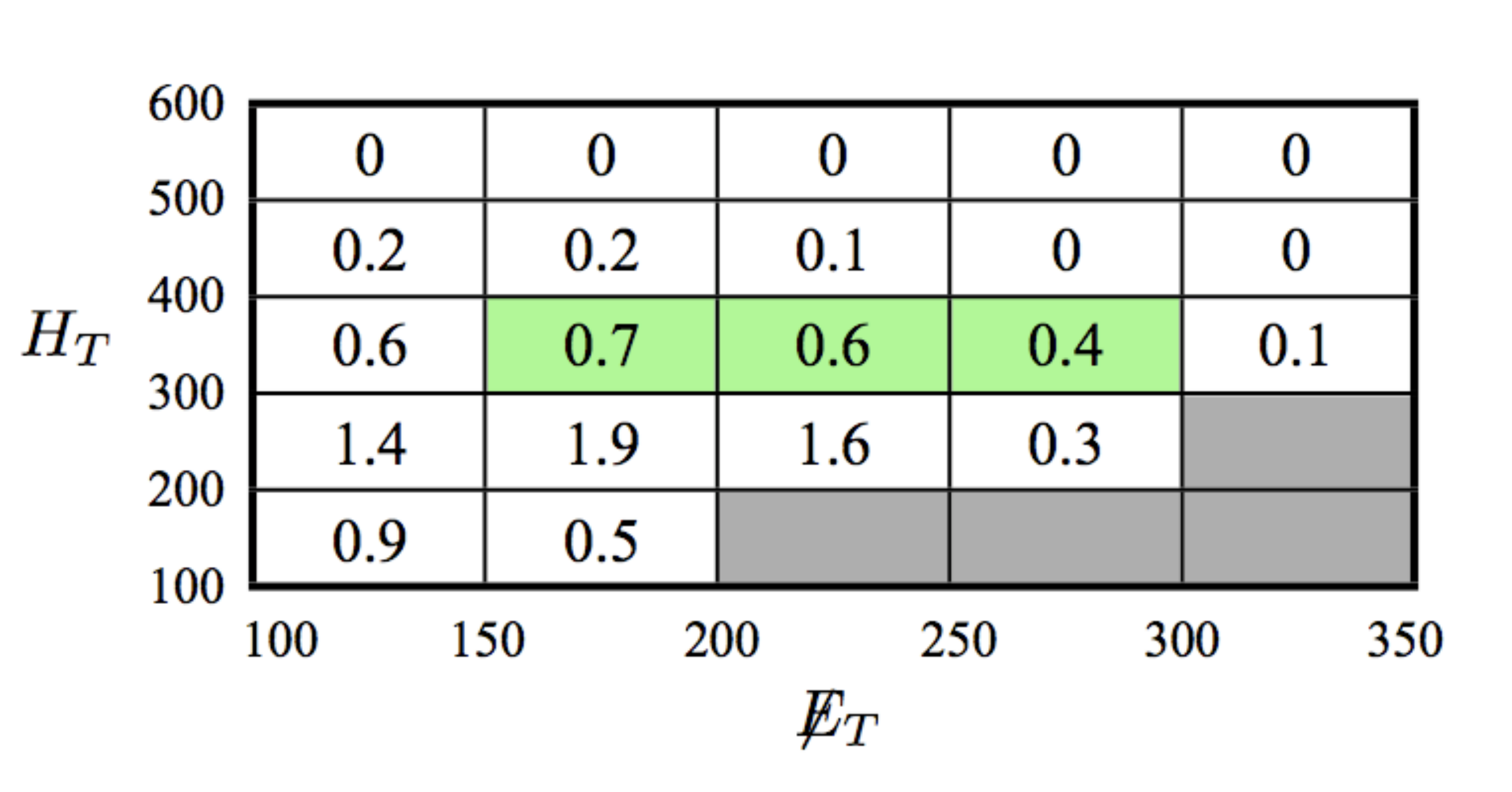}\\
    \hline
 \rotatebox{90}{\hspace{0.38in}Threejet} &
     \includegraphics[width=2.5in]{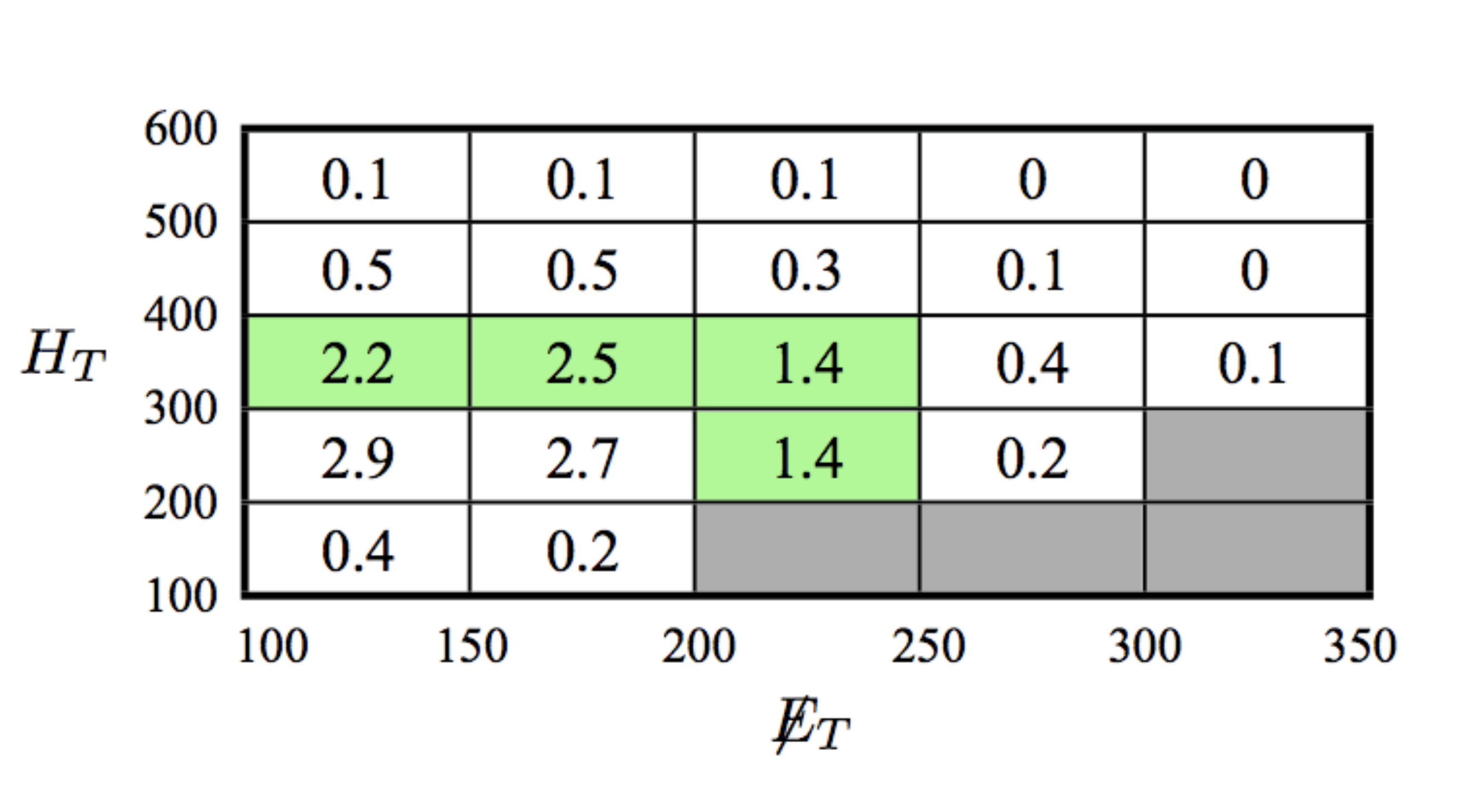}&

 \rotatebox{90}{\hspace{0.4in}Multijet}&
         \includegraphics[width=2.5in]{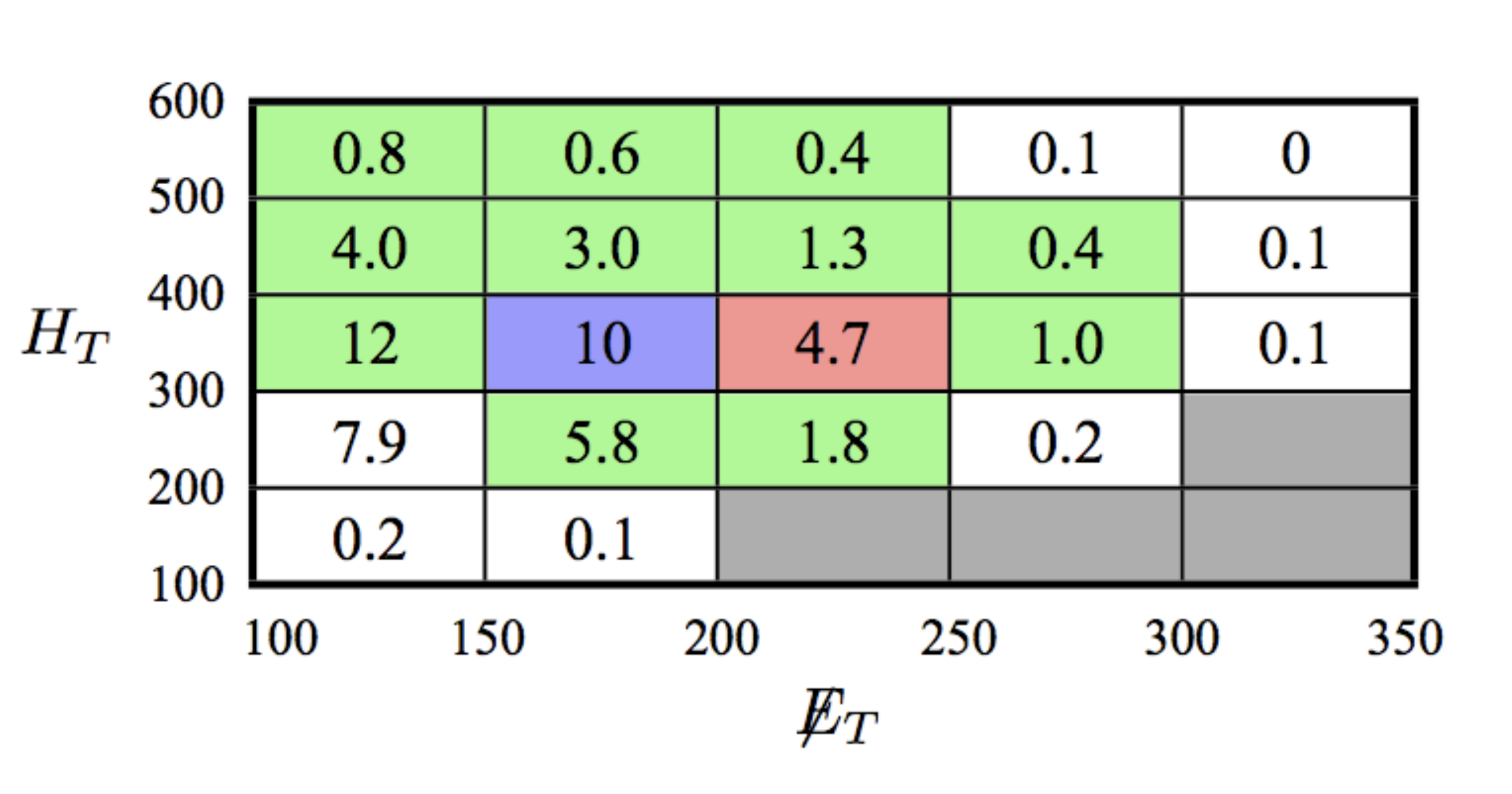}\\
  \hline
\end{tabular}
   \caption{Differential cross section (in fb) for the monojet, dijet, threejet, and multijet samples of a theoretical model spectrum with a 340 GeV gluino decaying directly into a 100 GeV bino (4 fb$^{-1}$).  Some boxes show significant deviation from the signal limits shown in Table II: green indicates $0.5 < \chi_i \leq 2$, blue indicates $2 < \chi_i \leq 3$, and red indicates $ \chi_i>3 $.  All boxes with $\chi_i > 1/2$ are included in the calculation of the total $\chi^2$ value.}
      \label{Fig: totalsignal}
\end{table}
\subsection{No Cascade Decays}
For the remainder of the paper, we will discuss how model-independent jets + $\MET$ searches can be used to set limits on the parameters in a particular theory.  We will focus specifically on the case of pair-produced gluinos at the Tevatron and begin by considering the simplified scenario of a direct decay to the bino.  The expected number of jets depends on the relative mass difference between the gluino and bino.  When the mass difference is small, the decay jets are very soft and initial-state radiation is important; in this limit, the monojet search is best.  When the mass difference is large, the decay jets are hard and well-defined, so the multijet search is most effective.  The dijet and threejet searches are important in the transition between these two limits.

\begin{figure}[b] 
   \centering
   \includegraphics[width=4.5in]{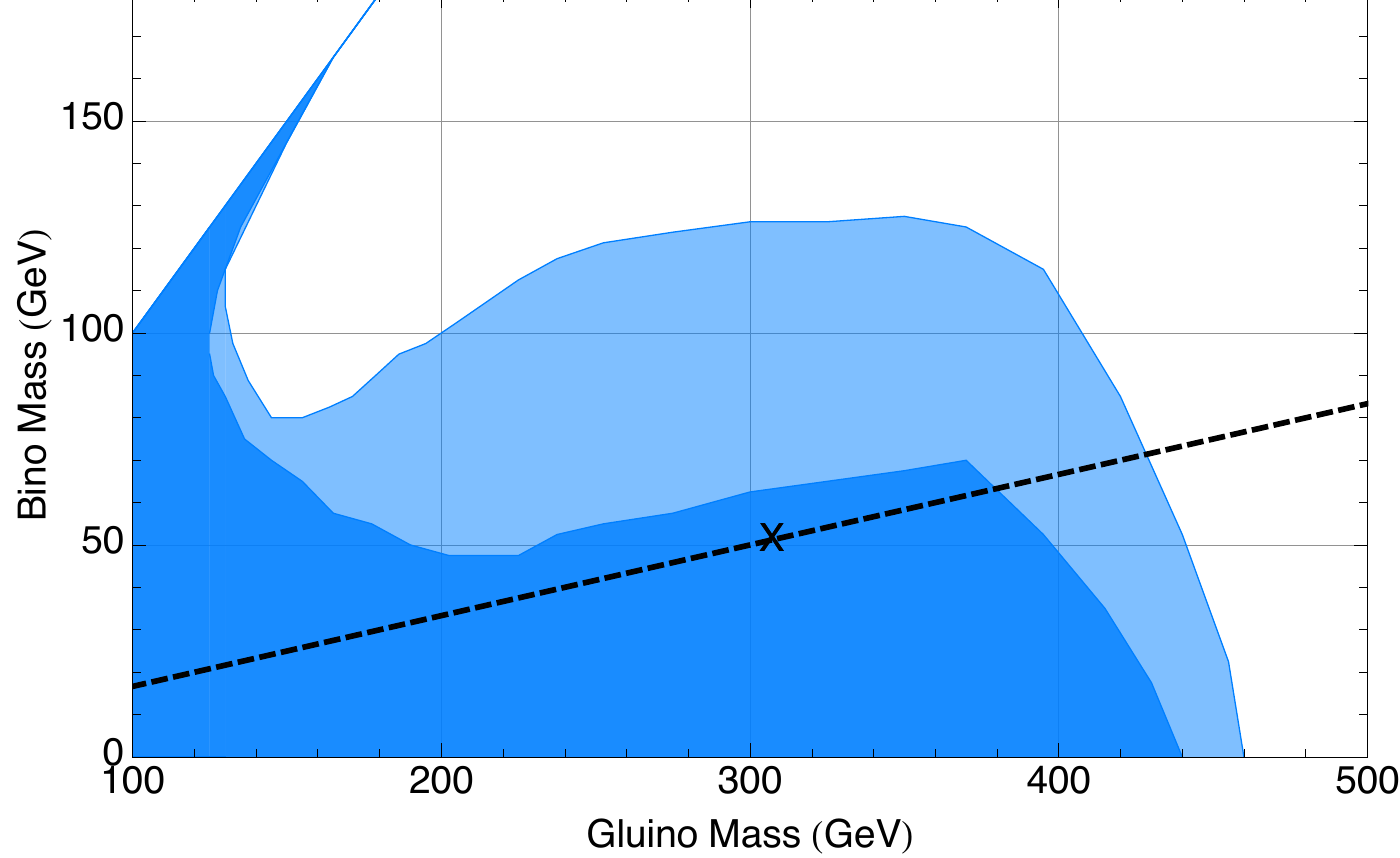}
   \caption{The 95\% exclusion region for $\DO$ at 4 fb$^{-1}$ assuming 50\% systematic error on background.  The exclusion region for a directly decaying gluino is shown in light blue; the worst case scenario for the cascade decay is shown in dark blue.  The dashed line represents the CMSSM points and the ``X" is the current $\DO$ exclusion limit at 2 fb$^{-1}$. }
   \label{Fig: decaycurve}
\end{figure}

As an example, let us consider the model spectrum with a 340 GeV gluino decaying directly into a 100 GeV bino.  In this case, the gluino is heavy and its mass difference with the bino is relatively large, so we expect the multijet search to be most effective.  Table \ref{Fig: totalsignal} shows the differential cross section grids for the 1-$4^+$ jet searches for this simulated signal point.  The colors indicate the significance of the signal over the limits presented in Table II; the multijet search has the strongest excesses.

Previously \cite{Alwall:2008ve}, we obtained exclusion limits by optimizing the $\MET$ and $H_T$ cuts, which involves simulating each mass point beforehand to determine which cuts are most appropriate.  This is effectively like dealing with a $1 \times 1$ grid, for which a 95\% exclusion corresponds to $\chi^2 = 4$.  The approach considered here considers the significance of all such cuts, and only requires that a single $n \times n$ differential cross section grid be produced for each search.   



Fig. \ref{Fig: decaycurve} shows the 95\% exclusion limit for directly decaying gluinos at 4 fb$^{-1}$ luminosity and 50\% systematic uncertainty on the background.  The results show that such gluinos are completely excluded for masses below $\sim130\GeV$.     


\subsection{Cascade Decays}

\label{Sec: Cascades}

\begin{figure}[t,b] 
   \centering
   \includegraphics[width=4.5in]{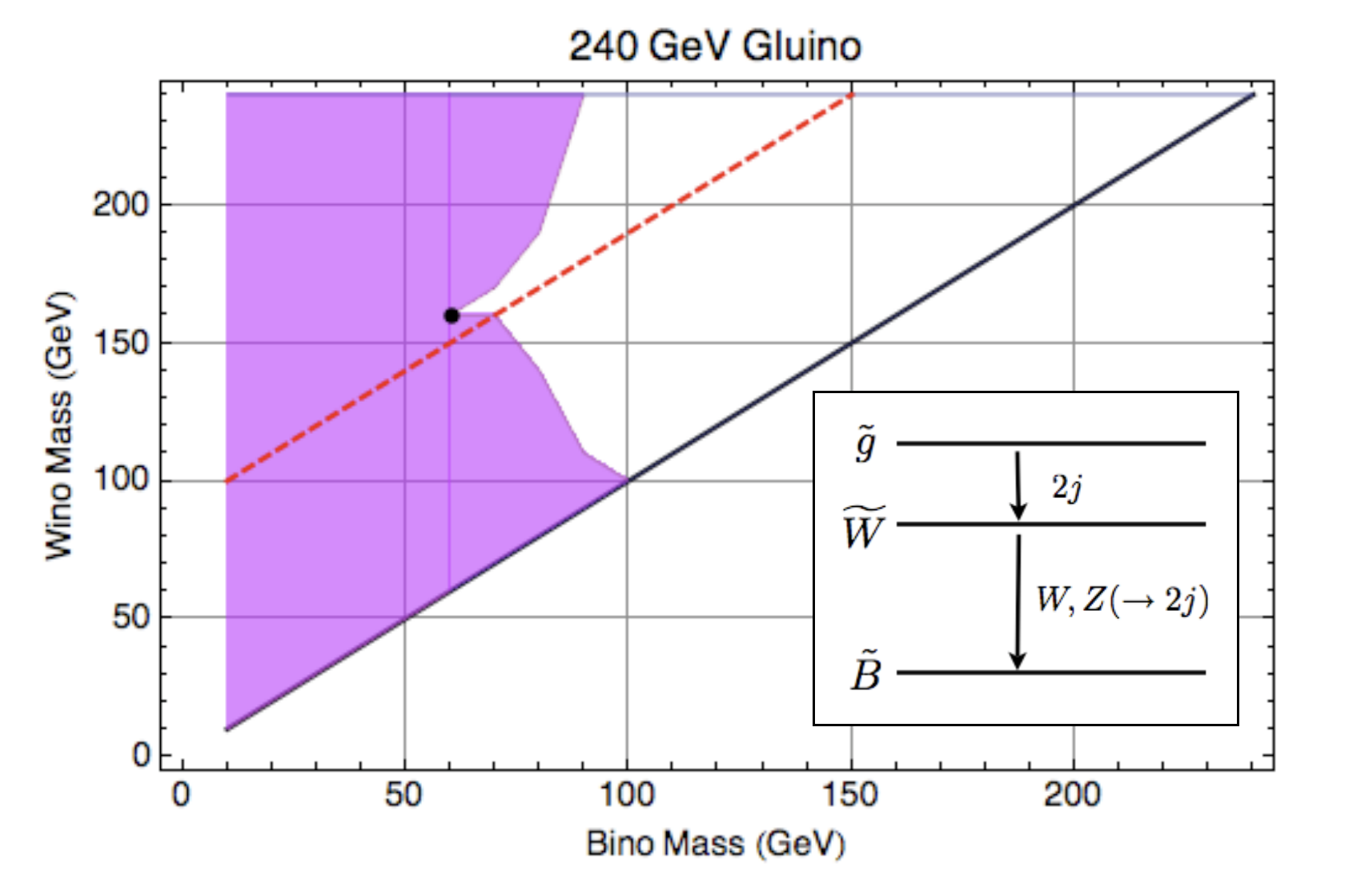}
   \caption{95\% exclusion region (purple) for a 240 $\GeV$ gluino decaying into a bino through a wino.  The dashed line is $m_{\widetilde{W}} = m_{\widetilde{B}} + \OO(m_{Z^0})$.  The black dot at $(m_{\widetilde{B}},m_{\widetilde{W}})$ = $(60,160)$, is the minimum bino mass for which a 240 GeV gluino is excluded for \emph{all} wino masses.  The inset shows the one-step cascade considered in the paper.}
   \label{Fig: fullscan}
\end{figure}

In this section, we will discuss the exclusion limits for the decay chain illustrated in the inset of Fig. \ref{Fig: fullscan}.  In general, cascade decays are more challenging to see because they convert missing energy to visible energy.\footnote{For additional discussion of model-independent searches of cascade decays at the Tevatron, see \cite{Weiner}.} 
The number of jets per event is greater for cascading gluinos than directly decaying ones and the spectrum of jet energies depends on the ratio of gaugino masses.  When $m_{\tilde{g}} \sim m_{\widetilde{W}}$, two hard jets are produced in the decay of the wino to the bino.  In the opposite limit, when $m_{\widetilde{W}} \sim m_{\widetilde{B}}$, two hard jets are produced in the decay of the gluino to the wino.  When $m_{\tilde{g}} < m_{\widetilde{W}} < m_{\widetilde{B}}$, four fairly hard jets are produced, diminishing the $\MET$ and making this region of parameter space the most challenging to see.  In particular, the most difficult region to detect is when
\begin{equation}
 m_{\widetilde{W}} = m_{\widetilde{B}} + \OO(m_{Z^0}).
 \label{Eq: crossover}
 \end{equation}  
In the region of parameter space, where $m_{\widetilde{W}} \sim m_{\widetilde{B}}$, the jets from the wino to bino decay become harder as the gauge bosons go on-shell. 

Fig. \ref{Fig: fullscan} shows the values of $m_{\widetilde{W}}$ and $m_{\widetilde{B}}$ that are excluded up to 95\% confidence for a $240\GeV$ gluino (shaded region).  The dark black dot, which represents the minimum bino mass for which a 240$\GeV$ gluino is excluded for all wino masses, falls close to Eq. \ref{Eq: crossover} (the dotted red line).

The exclusion region in Fig. \ref{Fig: fullscan} is not symmetric about the line $m_{\widetilde{W}} = m_{\widetilde{B}} + \OO(m_{Z^0})$.  
The asymmetry is a result of the hard lepton cuts.  When the gluino and wino are nearly degenerate, the leptons from the gauge boson decays are energetic, and these events are eliminated by the tight lepton cuts, reducing the significance below the confidence limit.  In the opposite limit, when the wino and bino are nearly degenerate, much less energy is transfered to the leptons and fewer signal events are cut.  Additionally, the jets produced in this case are color octets and give rise to a greater number of soft jets, as compared to the singlet jets emitted in the gauge boson decays.  The presence of many soft jets may decrease the lepton detection efficiency; as a result, it may be that even fewer events than expected are being cut.  

Figure \ref{Fig: decaycurve} compares the 95\% exclusion region for the cascade decay with that for the direct decay case.  The ``worst-possible" cascade scenario is plotted; that is, it is the maximum bino mass for which all wino masses are excluded.  For the one-step cascade considered here, gluinos are completely excluded up to masses of $\sim 125\GeV$.  


\subsection{$t$-channel squarks}
\label{Sec: t-channel}

\begin{figure}[t,b] 
   \centering
      \includegraphics[width=4.5in]{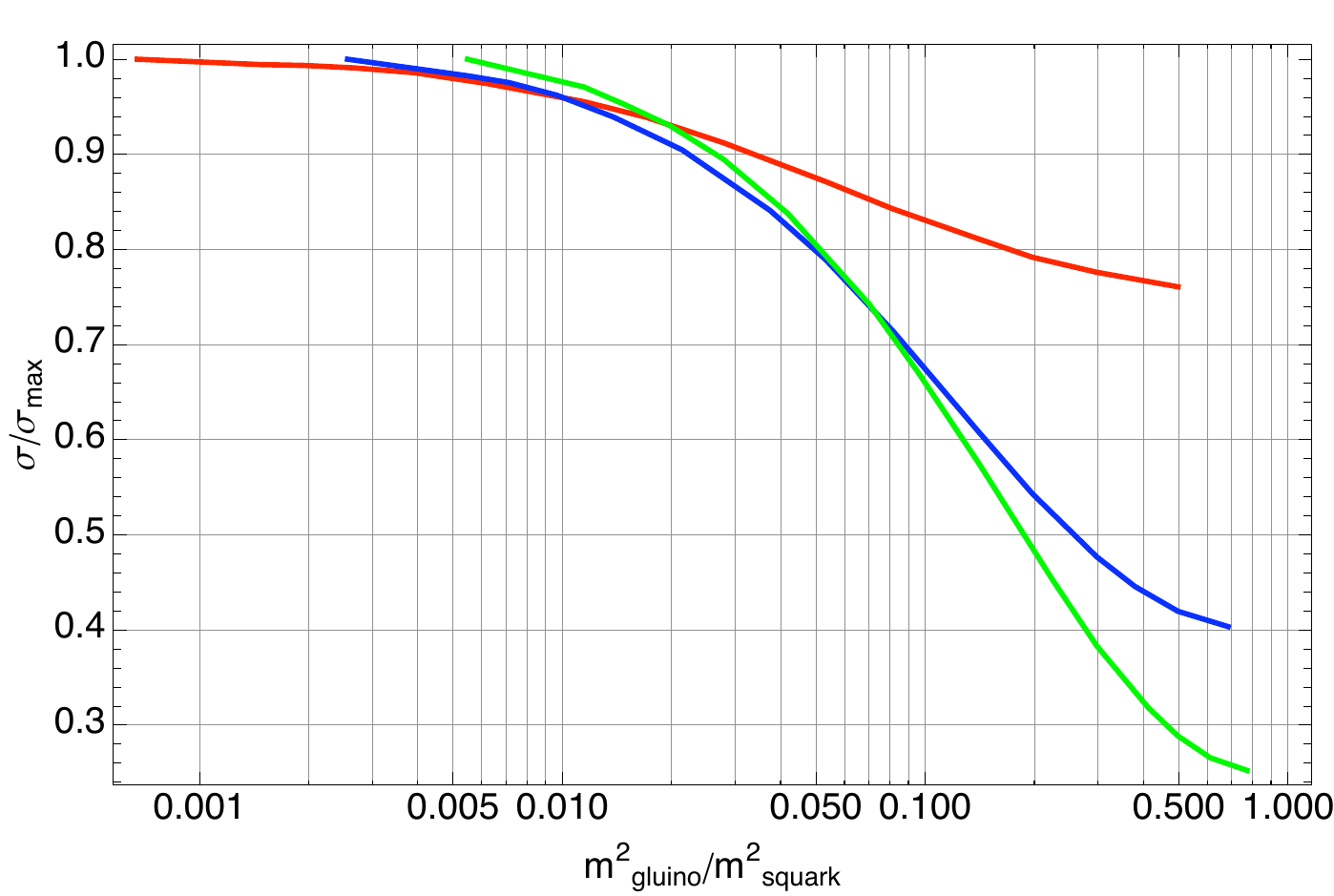}
   \caption{Gluino production cross section as a function of squark mass: (red) $m_{\tilde{g}} = 120$ GeV, (blue) $m_{\tilde{g}} = 240$ GeV, and (green) $m_{\tilde{g}} = 360$ GeV.}
   \label{Fig: squarks}
\end{figure}
Thus far, it has been assumed that the squarks are heavy enough that they do not affect the production cross section of gluinos.   If the squarks are not completely decoupled, they can contribute to $t$-channel diagrams in gluino pair-production.  Figure~\ref{Fig: squarks} shows the production cross section for a 120 GeV (red), 240 GeV (blue), and 360 GeV (green) gluino, as a function of squark mass.  When only one squark is light (and all the others are $\sim 4-5$ TeV), the production cross section is unaffected.  However, when the squark masses are brought down close to the gluino mass, the production cross section decreases by as much as $\sim$ 25\%, 60\%, and 75\% for $120$, $240,$ and $360\GeV$ gluinos, respectively.  A reduction in the production cross section alters the exclusion region in the gluino-bino mass plane; while the overall shape of the exclusion region remains the same, its size scales with the production cross section \cite{Alwall:2008ve}.

It is worthwhile to note, however, that while the inclusion of squarks reduces the exclusion region for pair-produced gluinos by decreasing the production cross section, it also provides alternate discovery channels through $\tilde{g} \tilde{q}$ or $\tilde{q} \tilde{q}$ production.  For example, if a gluino and squark are produced, with the gluino nearly degenerate with the bino, the subsequent decay of the squark will produce more visible energy than the gluino decay, thereby making the event more visible. 

\subsection{Monophoton Search}
\label{Sec: monophoton}

Initial-state QCD radiation is important for gaining sensitivity to degenerate gluinos.  Here, we will consider whether initial-state photon radiation may also be useful in the degenerate limit.  Such events are characterized by small $\MET$ and a hard photon.  

The main benefit of the monophoton search is that the Standard Model backgrounds are better understood; unlike the monojet case, QCD is no longer an important background.  Instead, the primary backgrounds come from processes such as $Z^0 (\rightarrow \nu \nu) + \gamma$, which is irreducible, and $W^{\pm} \rightarrow e^{\pm} \nu$ where the electron is mistaken as a photon or $W^{\pm} (\rightarrow l^{\pm} \nu) + \gamma$, where the lepton is not detected.  Other backgrounds may come from $W^{\pm}/Z^0$ + jet, where the jet is misidentified as a photon, or situations where muons or cosmic rays produce hard photons in the detector.

The $\DO$ Collaboration recently published results for their monophoton study, which searched for a Kaluza-Klein graviton produced along with a photon \cite{Abazov:2008kp}.  To reduce the Standard Model background, they required all events to have one photon with $p_T > 90\GeV$ and $\MET > 70\GeV$.  Events with muons or jets with $p_T > 15\GeV$ were rejected.  They estimate the total number of background events to be $22.4 \pm 2.5$.  

To investigate the sensitivity of monophoton searches to degenerate spectra, we consider several points and compared them against $\DO$'s background measurements.  We considered several benchmark values for gluino and bino masses and did a simple cuts-based comparison between the monophoton search and an optimized monojet search.  For example, Figure \ref{Fig: fullscan} shows that the monojet search safely excludes the case of a $140\GeV$ gluino and $130\GeV$ bino.  A monophoton search (with the cuts used in the $\DO$ analysis) gives $S/B = 0.48$ and $S/\sqrt{B} = 2.3$ for this mass point; thus the monophoton search is roughly as sensitive but has a lower $S/B$ value.  Similarly, a $120\GeV$ gluino and $100\GeV$ bino is safely excluded by the monojet search, but the monophoton search only gives $S/B = 0.39$ and $S/\sqrt{B} = 1.86$.  

There are several reasons why the monophoton search is not as successful as the monojet one.  In the degenerate gluino region, the possibility of getting jets with a  $p_T$ above the $15\GeV$ threshold is significant (even though the mass difference is $\OO(10\GeV)$) because the gluinos are boosted.  The monophoton search vetoes many events with such boosted decay jets.  In addition, getting photon ISR is much more difficult than getting QCD ISR for several reasons - most importantly, because $\alpha_{\text{EM}} \ll \alpha_s$ and because one is insensitive to the gluon-induced processes that contribute to the cross section.  
Despite these challenges,  the significance of the monophoton search could still increase sensitivity.    The monophoton does not
fare significantly more poorly than the monojet one with the current set of cuts.  Thus, it is possible that a more optimal set of cuts may increase the effectiveness of the search, especially given that the backgrounds are better understood in this case.
Finally, the above estimates do not account for the photon detection rate in PGS, which may be different from that used by $\DO$'s full detector simulator, from which the background estimates were taken.

\subsection{Leptons}
\label{Sec: leptons}
In this section, we address whether leptons from cascades can be used to augment the sensitivity of jets + $\MET$ searches.  In the gluino cascade decay considered in this paper, it is possible to get leptons from the $W^{\pm}$ and $Z^0$ boson decays.  The $10\GeV$ lepton veto, however, eliminates most of these events. 
The exclusion limit for the gluino decay discussed in Sec. \ref{Sec: Cascades} is not improved by removing the lepton veto; most of the irreducible backgrounds ($W^{\pm} + nj$ and $t\bar{t} + nj$) have a lepton and dominate over the signal when the veto is removed.  The exclusion limit is not improved even if we require all events to have a certain number of leptons, or place cuts on lepton $p_T$.

The question still remains as to whether there is any region in parameter space where the jets + $\MET$ study places no exclusion, but a jets + $\MET$ + lepton study does.  The lepton signal is useful for light gluinos ($\lesssim 250\GeV$) that are nearly degenerate with the wino.  The signal point, a $210\GeV$ gluino decaying to a $50\GeV$ bino through a $170\GeV$ wino, is not excluded by the ordinary jets + $\MET$ analysis.  We find here, though, that it has a significance\footnote{Here, the estimate of the significance only accounts for the statistical error; it does not include the systematic uncertainty.} of $\simeq 4.4$ for a $p_T$ cut of $50\GeV$, but with a $S/B \simeq 0.15$. 

For high-mass gluinos, inclusion of the lepton signal does not increase the sensitivity of the search because the smaller production cross section decreases the signal significance.  It might however be possible that lepton signatures are effective for high-mass gluinos in lepton-rich cascades that contain sleptons.  Overall, though, these results indicate that while jets + $\MET$ + lepton searches may be useful in certain regions of parameter space, they should be combined with jets + $\MET$ searches to provide optimal coverage.

\section{Conclusion}
\label{Sec: conclusion}

In this article, we discuss how model-independent bounds can be placed on the mass of the lightest color octet particle that is pair-produced at the Tevatron.  The main aspects of the analysis focus on the advantage of running exclusive $1j - 4^+j$ searches, and placing limits using the measured differential cross section as a function of the visible and missing energy.  We show that the exclusion reach can be significantly extended beyond those published by $\DO$ because the $\MET$ and $H_T$ cuts used in their analysis were only optimized for points in CMSSM parameter space.  The proposed analysis we present here opens up the searches to all regions of parameter-space, allowing us to set limits on all kinematically-accessible gluinos.  We also show how the exclusion reach is degraded when gluino cascade decays are included, focusing on the example of an intermediate wino, which decays to the dark matter candidate.  

We have so far only focused on jet classification, $\MET$, and $H_T$ as available handles for increasing the reach of jets + $\MET$ searches.  However, in certain special cases, other techniques might be useful.  For example, if the gluino decays dominantly to $b$ jets, heavy flavor tagging can be used advantageously.  

In our analysis of the cascade decays, we often found that the regions of highest significance in the differential cross section plot were pressed down against the 100 GeV cutoff in missing transverse energy.  This lower limit was imposed to avoid regions where the QCD background dominates.  If the 100 GeV limit could be reduced, then it would open up regions of high statistical significance that renders sensitivity to a larger region of parameter space.  The numerous uncertainties in the theory and numerical generation of QCD events make it unlikely that precision QCD background will be generated in the near future.  However, it may still be possible to reduce the cutoff by using event shape variables (i.e., sphericity).  

Looking forward to the LHC, jets + $\MET$ searches are still promising discovery channels for new physics.  The general analysis presented in this paper can be taken forward to the LHC without any significant changes.  
 The primary modification will be to optimize the jet $E_T$ used in the classification of the $nj+\MET$ searches.  The backgrounds for the LHC are dominantly the same; however $t\bar{t}$ will be significantly larger and the size of the QCD background will also be different.  Many of the existing proposals for searches at the LHC focus primarily on $4^+j+\MET$ inclusive searches and are insensitive to compressed spectra; see \cite{Kawagoe:2006sm} for further discussion on MSSM-specific compressed spectra at the LHC.   By having exclusive searches over $1j+\MET$ to $4^+j+\MET$, the LHC will be sensitive to most beyond the Standard Model spectra that have viable dark matter candidates that appear in the decays of new strongly-produced particles, regardless of the spectrum.  Additionally, having the differential cross section measurements will be useful in fitting models to any discoveries.   Finally, it is necessary to confirm that there are no gaps in coverage between the LHC and Tevatron; in particular, if there is a light ($\sim$125 GeV) gluino, finding signal-poor control regions  to measure the QCD background may be challenging.

\section*{Acknowledgements}
We would like to thank  Jean-Francois Grivaz, Andy Haas,  Frank Petriello, and Patrice Verdier and the CDF exotics group for helpful discussions.   JA, M-PL, ML, and JGW are supported by the DOE under contract DE-AC03-76SF00515 and partially by the NSF under grant PHY-0244728.  JA is supported by the Swedish Research Council.  ML is supported by NDSEG and Soros fellowships.

\providecommand{\href}[2]{#2}\begingroup\raggedright

\endgroup

\end{document}